\documentclass[fleqn,usenatbib]{mnras}

\usepackage{newtxtext,newtxmath}
\usepackage[T1]{fontenc}
\usepackage{ae,aecompl}

\usepackage{graphicx}	% Including figure files
\usepackage{amsmath}	% Advanced maths commands
\usepackage{color} % FJ 
\usepackage{slashbox,multirow,booktabs} % FJ, for tables
\usepackage{enumitem} % FJ, for itemization without bullets and indents
\usepackage[flushleft]{threeparttable} % FJ, for note under tables

\newcommand{\eq}[1]{eq.~(\ref{eq:#1})}

\newcommand{\se}[1]{\S\ref{sec:#1}}
\newcommand{\app}[1]{Appendix \ref{app:#1}}
\newcommand{\fig}[1]{Fig.~\ref{fig:#1}}

\newcommand{\tab}[1]{Table~\ref{tab:#1}}
\newcommand{\be}{\begin{equation}}
\newcommand{\ee}{\end{equation}}
\newcommand{\bad}{\begin{equation} \begin{aligned}}
\newcommand{\ead}{\end{aligned} \end{equation}}

\newcommand{\Msun}{M_\odot}

\newcommand{\kpc}{\,{\rm kpc}}
\newcommand{\pc}{\,{\rm pc}}
\newcommand{\Gyr}{\,{\rm Gyr}}
\newcommand{\cm}{\,{\rm cm}}
\newcommand{\g}{\,{\rm g}}

\newcommand{\rhoc}{\rho_{\rm crit}}

\newcommand{\rhodm}{\rho_{\rm dm}}
\newcommand{\rhodmc}{\rho_{\rm dm0}}
\newcommand{\rhob}{\rho_{\rm b}}

\newcommand{\rhoiso}{\rho_{\rm iso}}
\newcommand{\rhocdm}{\rho_{\rm cdm}}
\newcommand{\rhos}{\rho_{\rm s}}
\newcommand{\rhot}{\tilde{\rho}}

\newcommand{\Mv}{M_{\rm vir}}

\newcommand{\Mb}{M_{\rm b}}
\newcommand{\Mi}{M_{\rm i}}
\newcommand{\Mdmf}{M_{\rm dm, f}}
\newcommand{\Mdm}{M_{\rm dm}}
\newcommand{\Miso}{M_{\rm iso}}
\newcommand{\Mcdm}{M_{\rm cdm}}
\newcommand{\Mt}{\tilde{M}}

\newcommand{\fb}{f_{\rm b}}
\newcommand{\Rv}{R_{\rm vir}}

\newcommand{\rhalf}{r_{\rm 1/2}}
\newcommand{\rs}{r_{\rm s}}

\newcommand{\rc}{r_{\rm c}}
\newcommand{\rf}{r_{\rm f}}
\newcommand{\rt}{\tilde{r}}
\newcommand{\Lt}{\tilde{L}}

\newcommand{\Vc}{V_{\rm circ}}

\newcommand{\Vvir}{V_{\rm vir}}

\newcommand{\vt}{\tilde{v}}

\newcommand{\sigmam}{\sigma_m}
\newcommand{\sigmatm}{\tilde{\sigma}_m}
\newcommand{\ttilde}{\tilde{t}}

\newcommand{\tage}{t_{\rm age}}

\newcommand{\rmd}{{\rm d}}

\newcommand{\coreNFW}{{\footnotesize CORE}NFW\,}

\begin{document}

%%%%%%%%%%%%%%%%%%%%%%%%%%%%%%%%%%%%%%%%%%%%%%%%%%

%%%%%%%%%%%%%%%%%%% TITLE PAGE %%%%%%%%%%%%%%%%%%%

% Title of the paper, and the short title which is used in the headers.
% Keep the title short and informative.
\title[SIDM haloes]{A semi-analytic study of self-interacting dark-matter haloes with baryons}

% The list of authors, and the short list which is used in the headers.
% If you need two or more lines of authors, add an extra line using \newauthor
\author[Jiang et al.]{Fangzhou Jiang$^{1, 2}$ \thanks{Troesh Scholar, \href{mailto:fzjiang@caltech.edu}{fzjiang@caltech.edu}, \href{mailto:fjiang@carnegiescience.edu}{fjiang@carnegiescience.edu}}, 
Andrew Benson$^{1}$,
Philip F. Hopkins$^{2}$,
Oren Slone$^{3, 4}$, 
Mariangela Lisanti$^{3,5}$, 
\newauthor
Manoj Kaplinghat$^{6}$, 
Annika H. G. Peter$^{ 7, 8, 9}$,
Zhichao Carton Zeng$^{7, 8}$, 
Xiaolong Du$^{1}$,
Shengqi Yang$^{1}$,
\newauthor
Xuejian Shen$^{2}$
\vspace*{8pt}
\\
% List of institutions
$^{1}$ Carnegie Observatories, 813 Santa Barbara Street, Pasadena, CA 91101, USA \\
$^{2}$ TAPIR, California Institute of Technology, Pasadena, CA 91125, USA\\
$^{3}$ Department of Physics, Princeton University, Princeton, NJ 08544, USA\\
$^{4}$ Center for Cosmology and Particle Physics, Department of Physics, New York University, New York, NY 10003, USA\\
$^{5}$ Center for Computational Astrophysics, Flatiron Institute, 162 Fifth Ave, New York, NY 10010, USA\\
$^{6}$ University of Irvine, Irvine, CA 92697, USA\\
$^{7}$ Department of Physics, The Ohio State University, 191 W. Woodruff Ave., Columbus OH 43210, USA\\
$^{8}$ Center for Cosmology and Astroparticle Physics, The Ohio State University, 191 W. Woodruff Ave., Columbus OH 43210, USA\\
$^{9}$ Department of Astronomy, The Ohio State University, 140 W. 18th Ave., Columbus OH 43210, USA\\
}

% These dates will be filled out by the publisher
\date{}

% Enter the current year, for the copyright statements etc.
%\pubyear{2022}

% Don't change these lines
\label{firstpage}
\pagerange{\pageref{firstpage}--\pageref{lastpage}}
\maketitle

\begin{abstract}
We combine the isothermal Jeans model and the model of adiabatic halo contraction into a simple semi-analytic procedure for computing the density profile of self-interacting dark-matter (SIDM) haloes with the gravitational influence from the inhabitant galaxies. 
We show that the model agrees well with cosmological SIDM simulations over the entire core-forming stage and up to the onset of gravothermal core-collapse.
Using this model, we show that the halo response to baryons is more diverse in SIDM than in CDM and depends sensitively on galaxy size, a desirable link in the context of the structural diversity of bright dwarf galaxies.
The fast speed of the method facilitates analyses that would be challenging for numerical simulations -- notably, 1) we quantify the SIDM halo response as functions of the baryonic properties, on a fine mesh grid spanned by the baryon-to-total-mass ratio, $\Mb/\Mv$, and galaxy compactness, $\rhalf/\Rv$; 2) we show with high statistical precision that for typical Milky-Way-like systems, the SIDM profiles are similar to their CDM counterparts; and 3) we delineate the regime of gravothermal core-collapse in the $\Mb/\Mv-\rhalf/\Rv$ space, for a given cross section and a given halo concentration. 
Finally, we compare the isothermal Jeans model with the more sophisticated gravothermal fluid model, and show that the former yields faster core formation and agrees better with cosmological simulations. 
We attribute the difference to whether the target CDM halo is used as a boundary condition or as the initial condition for the gravothermal evolution, and thus comment on possible future improvements of the fluid model. 
We have made our programs for the model publicly available at \href{https://github.com/JiangFangzhou/SIDM}{https://github.com/JiangFangzhou/SIDM}.
\end{abstract}


%%%%%%%%%%%%%%%%%%%%%%%%%%%%%%%%%%%%%%%%%%%%%%%%%%

%%%%%%%%%%%%%%%%% BODY OF PAPER %%%%%%%%%%%%%%%%%%

%---------------------------------------------------------------------------------------------------------
\section{Introduction}

% Why SIDM
Self-interacting dark matter (SIDM) provides appealing revisions on small scales to the standard $\Lambda$+Cold Dark Matter (CDM) paradigm of cosmic structure formation.
Elastic self-interactions of dark-matter particles transfer heat towards the central regions of dark-matter haloes, creating constant-density, isothermal cores \citep[e.g.,][]{Kochanek00,Colin02,Vogelsberger12, Peter13, Rocha13}.
This is a convenient way of explaining the dark-matter cores in some dwarf galaxies \citep[e.g.,][]{deBlok08,Oh15}, without breaking the large-scale success of the standard cosmology. 

Galaxy formation complicates this picture. 
Hydro-cosmological SIDM simulations, as well as idealized SIDM-only simulations with analytical disk potentials, have shown that the dark-matter density profiles can sometimes be equally cuspy or cuspier than their CDM counterparts \citep[e.g.,][]{Sameie21,Elbert18}. 
This implies that the response of SIDM haloes to the inhabitant galaxies are diverse and highly sensitive to certain baryonic details.
The sensitivity of the SIDM halo response to baryonic details could be advantageous for explaining the small scale puzzles \citep[e.g.,][]{Kamada17, Creasey17, Ren19, Kaplinghat20, Zentner22}.
In fact, there is now compelling observational evidence that the structures of bright dwarf galaxies are diverse, not only in terms of the central dark-matter density slope \citep[e.g.,][]{Relatores19,Shi21} but also straightforwardly in terms of the galaxy size, which ranges from $\sim0.5\kpc$ for compact ellipticals \citep[e.g.,][]{Chilingarian15} all the way to $\sim5\kpc$ for ultra-diffuse galaxies \citep[e.g.,][]{Koda15}. 
These two aspects of structural diversity may actually be highly correlated, at least in $\Lambda$CDM. 
For example, simulated ultra-diffuse galaxies tend to be hosted by cored dark-matter haloes \citep[e.g.,][]{Jiang19}, where supernovae-driven gas outflows puff up simultaneously the galaxies and the host haloes.

%Why semi-analytic
It is therefore interesting to revisit the correlation between galaxy size and host halo structure in the context of SIDM. Can we quantify the halo response to baryons in simple terms? Is it stronger or weaker than that in CDM? Which baryonic process is the most important for establishing the galaxy-SIDM-halo relation?   
To answer these questions, hydro-cosmological SIDM simulations have been developed, however, they must find a balance between sample size and numerical resolution: zoom-in hydro-cosmological SIDM simulations have so far been limited to a small sample of Milky-Way-like systems and dwarfs \citep[e.g.,][]{Sameie21, Shen21, Cruz21}, whereas large-box SIDM simulations \citep[e.g.,][]{Robertson19} which contain large statistical samples still lack the resolution for reliably resolving the innermost few kpc. 
In this work, we adopt a semi-analytic approach based on the isothermal Jeans model first introduced in \citet{Kaplinghat14,Kaplinghat16}.
This model solves the Jeans-Poisson equation for the profile of the SIDM isothermal core, given the dark-matter density and velocity dispersion at the centre as well as the baryonic distribution. 
A recent adaptation of this method has been shown to be remarkably accurate compared to large-box SIDM simulations \citep{Robertson21}. 
We improve this model by adding a prescription for adiabatic halo contraction \citep{Gnedin04}, thus making it more self-consistent in describing the baryonic effect. 

This integrated model takes a target CDM halo and baryonic potential as inputs. 
It computes the contracted CDM halo given the baryonic potential, and stitches an isothermal SIDM core to the CDM-like outskirt by minimizing their differences at the transition radius within which collisions are frequent.
As such, this model can quickly compute density profiles for SIDM haloes with inhabitant galaxies, and, as we show below, produce results that are remarkably similar to those from {\it zoom-in} hydro-cosmological simulations.
The speed of this semi-analytic approach enables investigations of SIDM halo response with high statistical precision and with long baselines of input parameters such as baryonic size and mass. 

%Layout of this work ...
This paper is organized as follows. In \se{model}, we recap the model ingredients and combine them into a workflow, summarized in \se{workflow}.
In \se{comparison}, we compare the model predictions to the results from zoom-in cosmological SIDM simulations, including both dark-matter-only setups and hydro-simulations. 
After demonstrating the accuracy of the model, we use it to study the halo response in \se{HaloResponse}, where we quantitatively relate the inner structure of the SIDM haloes to the compactness and mass fraction of the inhabitant galaxies, and show the importance of considering adiabatic halo contraction.
Finally, in \se{discussion}, we compare this model to the other one-dimensional method for SIDM haloes that is extensively studied in the literature -- the gravothermal fluid model (\se{gravothermal}), and we also study the facilitation of gravothermal core-collapse by the inhabitant galaxy, providing regions of core-collapse in the space spanned by galaxy mass fraction and galaxy compactness, as a function of the cross section and target halo concentration. 
For general readers who want to skip the technical details and get to the results sooner, \se{workflow} can be a good starting point. 

Throughout, we define the virial radius of a distinct halo as the radius within which the average density is $\Delta=200$ times the critical density for closure. 
We also assume spherical symmetry for both the dark-matter haloes and galaxies.
We adopt a flat cosmology with the present-day matter density $\Omega_{\rm m}=0.3$, baryonic density $\Omega_{\rm b}=0.0465$, dark energy density $\Omega_\Lambda=0.7$, a power spectrum normalization $\sigma_8=0.8$, a power-law spectral index of $n_s=1$, and a Hubble parameter of $h=0.7$, unless otherwise mentioned.

%---------------------------------------------------------------------------------------------------------
\section{Analytic method for computing the density profile of SIDM haloes}\label{sec:model}

Scattering between dark-matter particles is prevalent in the centre of a halo where the dark-matter density is high, but is infrequent on the outskirts where the scattering timescale is longer than the lifetime of the halo. 
The full profile of an SIDM halo therefore consists of a thermalized core and a CDM-like outer region.
The transition is around a characteristic radius $r_1$, within which an average dark-matter particle has experienced more than one scattering over the lifetime $\tage$ of the halo \citep{Kaplinghat16}:
\be\label{eq:r1}
\frac{4}{\sqrt{\pi}}\rhodm(r_1)v(r_1) \sigmam = \frac{1}{\tage},
\ee
where the left-hand side is the scattering rate per particle, with $\rhodm$ the DM density, $(4/\sqrt{\pi})v$ the average relative velocity between DM particles for a Maxwellian distribution (where $v$ is the 1D velocity dispersion), and $\sigmam$ the self-interaction cross-section per particle mass. 
Note that the cross section also carries a radius dependence if it is velocity dependent, which comes in via the velocity dispersion profile, i.e., $\sigma_m = \sigma_m[v(r)]$. 
Here, we assume constant cross section in the velocity-dispersion regime of interest. This assumption holds when the halo develops its isothermal core within $r_1$.

The impact of DM self-interactions on the halo density profile can be regarded as a modification to the inner part ($r<r_1$) of a CDM counterpart, and can be computed using the spherical Jeans equation with the assumption that the halo is isothermal within $r_1$ and in approximate equilibrium. 

%---------------------------------------
\subsection{Profile of the isothermal core}\label{sec:core}

The density profile of the isothermal dark-matter core can be solved by combining the spherical Jeans equation and the Poisson equation:
\be\label{eq:Jeans}
\frac{\rmd (\rhodm v^2)}{\rmd r}+\frac{2\beta}{r}v^2 = -\rhodm\frac{\rmd \Phi}{\rmd r},
\ee
\be\label{eq:Poisson}
\frac{1}{r^2} \frac{\rmd }{\rmd r} \left( r^2 \frac{\rmd\Phi} {\rmd r} \right)= 4\pi G \rho = 4\pi G(\rhodm+\rhob),
\ee
where $\Phi$ is the total gravitational potential,  $\rho$ is the total density, and $\rhob$ is the baryon density. 
With the assumption of an isotropic ($\beta=0$) and constant 1D velocity dispersion ($v(r)=v_0$), the Jeans equation has a simple generic solution:  
\be\label{eq:JeansSoln}
\rhodm(r) = \rhodmc \exp\left[-\frac{\Delta \Phi(r)}{v_0^2}\right] \,\,\, {\rm or}\,\,\, \Delta \Phi(r) = -v_0^2 \ln\left[\frac{\rhodm(r)}{\rhodmc}\right],
\ee
where $\rhodmc$ is the central dark-matter density, and $\Delta \Phi(r) = \Phi(r)-\Phi(0)$ is the potential difference between radius $r$ and the centre. 
Combining \eq{Poisson} and \eq{JeansSoln}, we get
\be \label{eq:JeansPoisson}
\frac{1}{r^2} \frac{\rmd }{\rmd r} \left( r^2 \frac{\rmd\ln\rhodm(r) } {\rmd r} \right) = -\frac{4\pi G}{v_0^2}[\rhodm(r) + \rhob(r)].
\ee
Following \citet{Kaplinghat14}, we assume a Hernquist profile for the baryon distribution,
\be
\rhob(r) = \frac{\Mb/2\pi r_0^3}{\frac{r}{r_0}\left(1+\frac{r}{r_0}\right)^3}, 
\ee
where $\Mb$ is the baryon mass, and $r_0$ the scale radius. 
Then, \eq{JeansPoisson} can be rewritten as the dimensionless form:
\be \label{eq:JeansPoissonSimplified}
\frac{\rmd^2h}{\rmd y^2} + \frac{2}{y}\frac{\rmd h}{\rmd y} + \frac{b}{y}+\frac{a e^h}{(1-y)^4}=0,
\ee
where $h(y) \equiv \Phi(y)/v_0^2$, $y = (r/r_0)/(1+r/r_0)$, $a\equiv4\pi Gr_0^2\rhodmc/v_0^2$, and $b\equiv2G\Mb/r_0v_0^2$. 
%Here, $b$ is a measure of the ratio between the baryon potential and the DM potential in the centre. 
The boundary conditions for solving this equation are $h(0)=0$ and $h^\prime(0)=-b/2$. 
The isothermal core profile can therefore be obtained by integrating \eq{JeansPoissonSimplified}, given the baryon properties ($\Mb$, $r_0$), the central DM density ($\rhodmc$), and the constant velocity dispersion within the core ($v_0$). 

There are four parameters in total that fully determine the isothermal dark-matter profile: two for baryons ($\Mb$, $r_0$) and two for dark matter ($\rhodmc$, $v_0$). 
For modelers, the baryonic parameters ($\Mb$, $r_0$) are usually known -- for constructing simple toy halo models based on observations, ($\Mb$, $r_0$) are available from surface photometry; for building more complex semi-analytic or semi-empirical frameworks, they can be set from empirical abundance-matching relations.
However, the DM parameters ($\rhodmc$, $v_0$) are not readily known. 
They need to be determined iteratively given the virial mass $\Mv$ and concentration $c$ of the target CDM halo, as we will describe in \se{stitching}.

We emphasize that, the isothermal Jeans model assumes that the system is in approximate equilibrium. 
Strictly speaking, an SIDM halo is never in Jeans equilibrium, but constantly evolving by transporting energy from the dynamically hotter region to colder places. 
For a target system that is initially described by a CDM profile, the dynamically hottest place is where the $v(r)$ profile peaks, so with self-interactions, the heat flows to the centre. 
As the system evolves, the core temperature gradually becomes the highest and then conducts energy outwards. The full time evolution can be described using the gravothermal fluid equations (see \se{gravothermal}).

%---------------------------------------
\subsection{Halo contraction} \label{sec:contraction}

The dark-matter distribution contracts in response to the condensation of baryons in the halo centre. 
\citet{Blumenthal86} described this process assuming circular orbits and an adiabatic invariant of $M(r)r$, where $M(r)$ is the total mass enclosed within radius $r$. 
\citet{Gnedin04} showed that the original adiabatic-contraction treatment overestimates the magnitude of contraction compared to the results of cosmological hydro-simulations, and attributed the mismatch to the oversimplified assumption of circular orbits. 
To account for orbital eccentricity and orbital phase distributions, they proposed a modified invariant, $M(\bar{r})r$, where $\bar{r}$ is the orbit-averaged radius for particles at instantaneous radius $r$, approximated by
\be
\bar{x} = A x^w, 
\ee
where $x=r/\Rv$, and the parameters $A\approx0.85$ and $w\approx0.8$ are calibrated with simulations. There is some halo-to-halo variation in these parameters \citep{Gnedin11}, which we ignore in this work.\footnote{We ignore the halo-to-halo variation because there seems to be no systematic trend of $w$ or $A$ with halo mass or concentration. $w$ is weakly dependent on the details of cooling, but usually within 0.6-1.0.}
With $M(\bar{r})r$ invariant and assuming that the baryons are initially distributed with the same radial profile as the dark matter, one can show that the final radius $\rf$ of dark-matter particles initially located at $r>\rf$ obeys the equation:
\be\label{eq:contraction}
\frac{r}{\rf} = 1-\fb + \frac{\Mb(\bar{\rf})}{\Mi(\bar{r})},
\ee
where $\fb=\Mb/\Mv$ is the galactic mass fraction within $\Rv$, $\Mb(r)$ is the final baryon mass within $r$, and $\Mi(r)$ is the initial total mass profile.

Assuming that the initial distribution of DM and baryons both follow an NFW profile \citep{NFW97}, 
\be
\label{eq:NFW}
\rho(r) = \frac{\rhos}{cx\left(1+cx\right)^2},\,\,\, {\rm where} \,\,\,  \rhos = \frac{c^3}{3f(c)}\Delta \rhoc, 
\ee
with $\Delta$ the average overdensity with respect to the critical density of the Universe $\rhoc(z)$, $c$ the concentration parameter, and $ f(c)=\ln(1+c) - c/(1+c)$,
and that the final baryonic distribution obeys a Hernquist profile, then a solution of \eq{contraction} can be obtained.
The details of this step can be found in the appendix of \citet{Gnedin04}.
Solving \eq{contraction} for $\rf$ for an initial radius $r$, we get the enclosed mass profile $\Mdmf(\rf) = (1-\fb)\Mi(r)$ of the contracted halo.

The contracted DM mass profile is non-parametric. 
To facilitate subsequent modeling, such as solving for the characteristic radius $r_1$, we need simple, parametric expressions for the density profile $\rhodm(r)$ and the velocity-dispersion profile $v(r)$. 
We therefore fit the profile of a contracted halo with the Dekel-Zhao (DZ) profile \citep{Freundlich20}, which has analytic expressions for $\rhodm(r)$ and $v(r)$, and is flexible enough in the centre to account for the contraction, at the expense of adding just one more degree of freedom than NFW. 
The enclosed mass of a DZ profile is given by 
\bad\label{eq:DZmass}
\Mdm(r) = (1-\fb)\Mv\frac{g(cx,\alpha)}{g(c,\alpha)},
\ead
where $g(\xi,\alpha) = [\xi^{1/2}/(1+\xi^{1/2})]^{2(3-\alpha)}$; and $c$ and $\alpha$ are the free parameters describing the concentration and innermost density slope of the halo.
The density profile and the velocity dispersion profile are given by
\bad\label{eq:DekelDensity}
\rhodm(r) = \frac{\rho_{\rm DZ}}{x^\alpha (1+x^{1/2})^{2(3.5-\alpha)}}, \\
\ead
\be\label{equ:DekelVelocityDispersion}
v^2(r) = 2\Vvir^2 \frac{c}{g(c,\alpha)} \frac{x^{3.5}}{\chi^{2(3.5-\alpha)}} \sum_{i=0}^{8} \frac{(-1)^i 8!}{i!(8-i)!}  \frac{1-\chi^{4(1-\alpha)+i}}{4(1-\alpha)+i},
\ee
where $\rho_{\rm DZ} =[c^3(3-\alpha)]/[3g(c,\alpha)]\times\Delta \rhoc$,  $\Vvir$ is the circular velocity at the virial radius, and $\chi=x^{1/2}/(1+x^{1/2})$.
We fit the mass profile $\Mdmf(\rf)$ of a contracted halo using \eq{DZmass} and then solve \eq{r1} for the transition radius $r_1$ using the density and velocity dispersion of the best-fit DZ profile.
For typical baryon distributions ($0.01\la\fb\la0.2$ and $0.005\la r_0/\Rv\la 0.1$), the best-fit DZ profile agrees with the non-parametric solution of $\Mdmf(\rf)$ to per-cent level.

From now on, we drop the `dm' in the subscription of the symbol for central DM density $\rhodmc$ and simply denote it by $\rho_0$.

%---------------------------------------
\subsection{Stitching the isothermal core to the CDM outskirt}\label{sec:stitching}

\begin{figure*}
	\includegraphics[width=\textwidth]{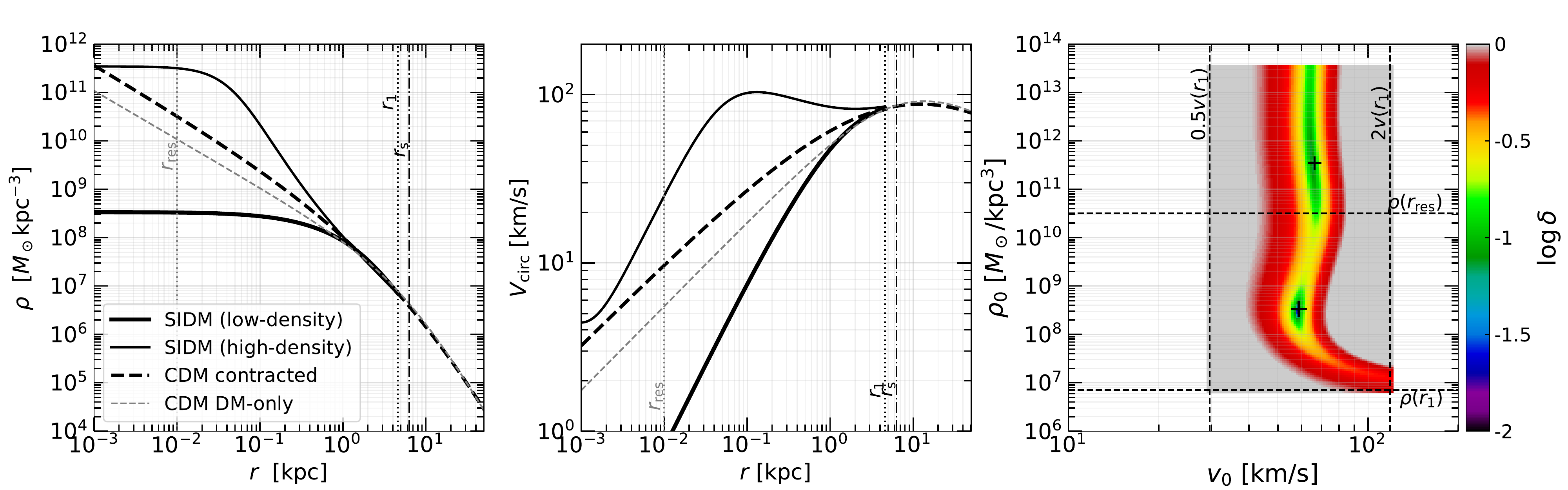}
    \caption{Illustration of the semi-analytical workflow (\se{workflow}) -- example of finding the SIDM profile for a cross-section of $\sigmam=1\cm^2/\g$ and a target CDM halo formed $\tage=5\Gyr$ ago with a present-day virial mass of $\Mv=10^{11}\Msun$ and a concentration of $c=15$. 
    The inhabitant galaxy has a total cold baryon mass of $\Mb=10^{9}\Msun$, and half-mass radius of $\rhalf=1.9\kpc$ ($\rhalf/\Rv=0.02$).  
    The thin grey dashed lines show the density profile ({\it left}) and circular-velocity profile ({\it middle}) of the original dark-matter-only target CDM halo; and the thick dashed lines are the profiles for the adiabatically contracted CDM halo.
    The thicker solid black lines are the profiles of the best-matching SIDM isothermal core, which corresponds to the low-density solution of $(\rho_0,v_0)$ as marked by the bigger black `+' sign in the right-hand panel.
    The thinner solid black lines are the profiles that correspond to the {\it discarded} high-density solution, as marked by the smaller `+' sign in the right-hand panel.
    The right-hand panel shows the colour map of the `stitching error' $\delta$, defined in \eq{StitchingError}, in the space of central density $\rho_0$ versus central velocity dispersion $v_0$. 
    Clearly there are two $\delta$ minima, but only the low-density solution agrees with simulation results (see \fig{SIDMprofilesPippin}). 
    As shown in Appendix \ref{app:TwoSolutions}, the two solutions get closer as the system evolves (i.e., as $\tage$ or cross-section increases). 
    When they join, gravothermal core-collapse starts to speed up (see \se{GC}). 
    The vertical and horizontal dashed lines indicate the region $\rhocdm(r_1)<\rho_0<\rhocdm(r_{\rm res}=10\pc)$ and $0.5v(r_1)<v_0<2v(r_1)$ , which brackets the low-density solution uniquely for a wide range of halo age and cross section.
    }
    \label{fig:stitching}
\end{figure*}

To obtain the full profile of a SIDM halo with baryons, we determine the parameters ($\rho_0$, $v_0$) of the isothermal core iteratively, such that the core joins the contracted CDM halo at radius $r_1$ smoothly in terms of the local density and the enclosed mass. 
Specifically, we search the space of  $\rho_0$-$v_0$ to minimize the following objective quantity:
\be\label{eq:StitchingError}
\delta^2 = \left[\frac{\rhoiso(r_1)-\rhocdm(r_1)}{\rhocdm(r_1)} \right]^2 + \left[\frac{\Miso(r_1)-\Mcdm(r_1)}{\Mcdm(r_1)} \right]^2,
\ee
where $\rhoiso(r)$ and $\rhocdm(r)$ are the density profiles, and $\Miso(r)$ and $\Mcdm(r)$ are the enclosed DM mass profiles, of the isothermal core and the contracted CDM halo, respectively.
There are two minima of $\delta^2$ in the $\rho_0$-$v_0$ space, with similar $v_0$ values but very different $\rho_0$.
The existence of the two solutions was already noted by \citet{Elbert18}. 
Here we illustrate them clearly in the right-hand panel of \fig{stitching}.

\citeauthor{Elbert18} only accepted the lower-density solution as it agrees with their simulation results better.
We emphasize that both solutions are physical in the sense that they both meet the requirement of constant temperature below $r_1$. 
It is just that realistic haloes form with properties closer to the lower-density solution, which is why the lower-density solution agrees better with cosmological simulation results.  
We find by trial and error that a practical searching range for the lower-density solution is $\rhocdm(r_1)<\rho_0<\rhocdm(10\pc)$ and $0.5v(r_1)<v_0<2v(r_1)$, which, in most cases, brackets a unique minimum of $\delta^2$. 

As will be shown below, this simple formalism can capture the onset of gravothermal core-collapse. 
As the halo age $\tage$ increases or as the cross section $\sigmam$ becomes larger, the two minima of $\delta^2$ get closer -- they first both decrease in $\rho_0$; then the lower-density solution turns around, manifesting the onset of gravothermal core-collapse; and finally the two solutions merge as core-collapse speeds up, beyond which point the isothermal model is no longer applicable. 
This is illustrated in \app{TwoSolutions}, and the high-density solution is therefore also useful, as we will address further in \se{GC}. 

%---------------------------------------
\subsection{Workflow}\label{sec:workflow}

We summarize the workflow for getting the density profile of a SIDM halo with baryons as follows:
\begin{itemize}[leftmargin=*]
\item[1.]  Given a CDM halo described by an NFW profile (i.e., with known virial mass $\Mv$, concentration $c$, and age $\tage$), and given an inhabitant galaxy described by a Hernquist profile (parameterized by the mass $\Mb$ and scale size $r_0$), compute the adiabatically contracted halo profile (\se{contraction}).
\item[2.]  Given the self-interaction cross-section, $\sigma_m$, solve \eq{r1} for the radius of frequent scattering, $r_1$, using the density profile and velocity-dispersion profile of the contracted CDM halo. 
\item[3.]  Integrate the spherical Jeans-Poisson equation, \eq{JeansPoisson}, to obtain an isothermal core profile  (\se{core}) -- do this iteratively to find the central DM density $\rho_0$ and the central velocity dispersion $v_0$ by minimizing the relative stitching error defined in \eq{StitchingError} (\se{stitching}).
%There are two solutions of  ($\rho_0$, $v_0$) for each input of target CDM halo ($\Mv$, $c$) and target galaxy ($\Mb$, $r_0$), and we accept the lower-density solution which agrees with cosmological SIDM simulations. 
\end{itemize}
To illustrate, \fig{stitching} shows an example of the density and circular velocity profiles of an SIDM halo obtained with this workflow. 
In this example, we adopt a self-interaction cross-section of $\sigmam=1\cm^2/\g$ and a target CDM halo of $\Mv=10^{11}\Msun$, $c=15$, and $\tage=10\Gyr$ with a Hernquist baryon distribution of mass $\Mb=10^{9}\Msun$ and half-mass radius $\rhalf=1.9\kpc$ (i.e., a Hernquist $r_0 = \rhalf/(1+\sqrt{2}) \approx 0.8\kpc $). 
These choices are largely arbitrary for illustration purposes, but are of the same of order as the Large Magellanic Cloud (LMC). 
In \app{TwoSolutions}, we demonstrate how the two solutions evolve as the halo age increases, and discuss in \se{GC} that the high-density solution can help us to phenomenologically predict the onset of gravothermal core-collapse. 
While this procedure is devised for haloes with baryons, it is fully compatible with dark-matter-only cases, for which one simply sets $\Mb$ small and $\rhalf$ large.

%---------------------------------------------------------------------------------------------------------
\section{Comparison with cosmological SIDM simulations}\label{sec:comparison}

In this section, we show that the aforementioned workflow gives halo profiles closely matching those from cosmological SIDM simulations. 
We also provide a simple analytical fitting formula for the dark-matter-only cases.  

%---------------------------------------
\subsection{Comparison with dark-matter only simulations}\label{sec:Pippin}

\begin{figure}
\includegraphics[width=0.48\textwidth]{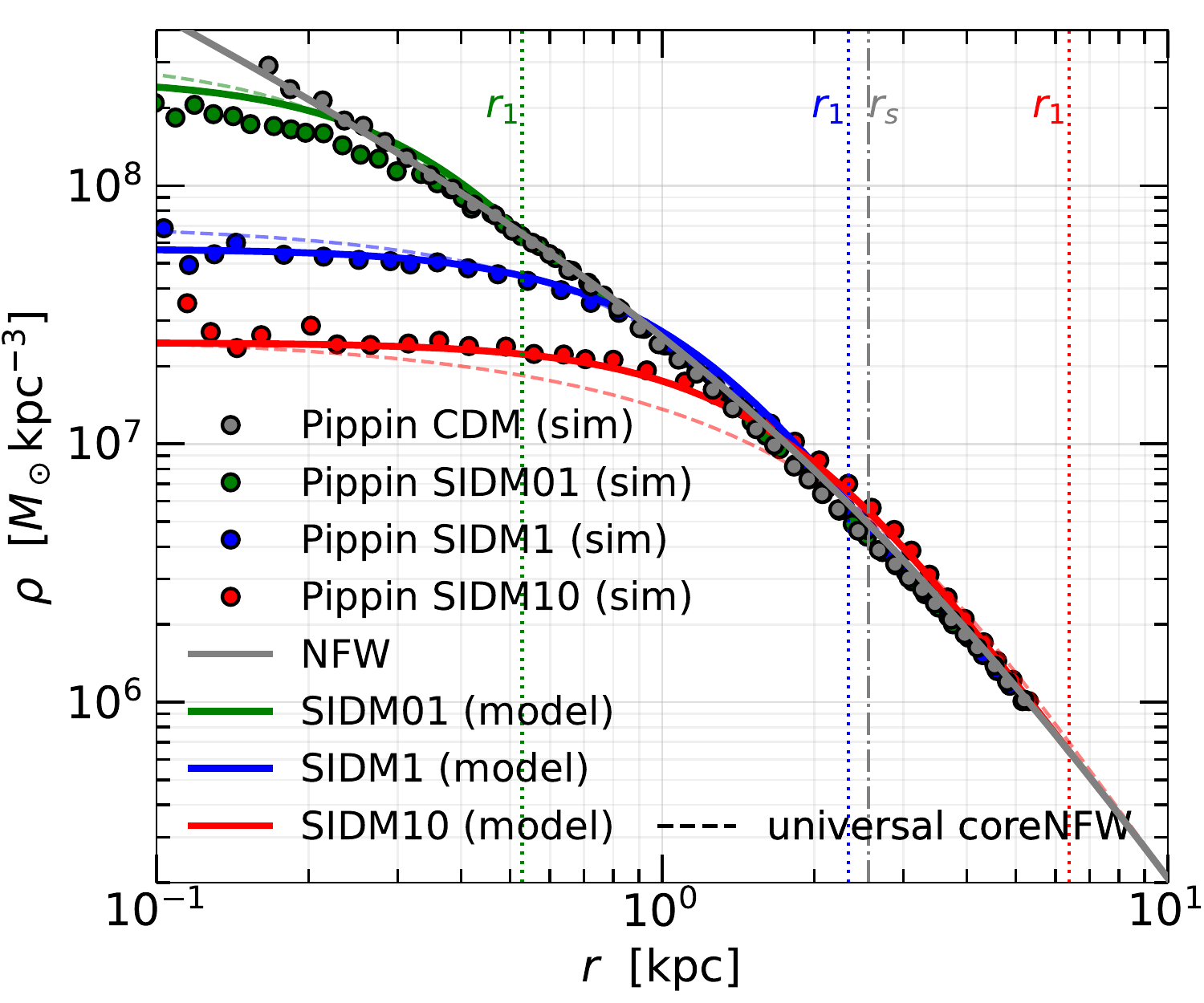}
\caption{ Comparison of the dark-matter density profiles from the model and from the cosmological $N$-body simulations of the Pippin haloes as in \citet{Elbert15} at $z=0$.  
The grey circles represent the reference-CDM simulation result; and the green, blue, and red circles represent the SIDM simulation results of cross-sections of $\sigma_m=0.1$, $1$, and $10$ $\rm cm^2/g$, respectively (labeled as SIDM01, SIDM1, and SIDM10).
The CDM halo is well described by an NFW profile of $\Mv=10^{9.89}\Msun$ and $c=15.8$, as indicated by the grey solid line -- this is used as the starting point of the isothermal Jeans model. 
The model predictions are shown by the solid lines of corresponding colours. 
The vertical dotted lines show the $r_1$ radii.
The model predictions agree very well with the simulation results across 2 dex in cross section. 
The thin dashed lines in pale colours represent a {\it universal} approximation, which is the coreNFW profile with a scale radius $\rc$ that is 0.45 times the respective $r_1$.}
\label{fig:SIDMprofilesPippin}
\end{figure}

To compare the model to cosmological dark-matter-only simulations, we use the zoom-in simulations of \citet{Elbert15} and focus on the `Pippin' haloes therein. 
The simulations adopt the {\it Wilkinson Microwave Anisotropy Probe-7} cosmology \citep{Komatsu11}, with $h=0.71, \Omega_{\rm m}=0.266, \Omega_{\rm \Lambda}=0.734, n_{\rm s}=0.963$ and $\sigma_8=0.801$.
For the high-resolution runs that we compare to, the particle mass is $1.5\times10^3 \Msun$, and the Plummer equivalent force softening length is 28 pc.
The Pippin halo was run in both CDM and SIDM with a wide range of velocity-independent cross-sections of $\sigma_m=0.1-50$  $\rm cm^2/g$, all starting from the same initial conditions. 
The SIDM implementation follows that of \citet{Rocha13}.
The CDM Pippin halo is accurately described by an NFW profile with a virial mass of $\Mv = 10^{9.89}\Msun$ and a concentration of $c=15.8$, as shown by the grey line in \fig{SIDMprofilesPippin}.
We use this NFW profile as the input of the target CDM profile for our model, and compute the SIDM profiles for $\sigma_m=0.1$, $1$, and $10\, \rm cm^2/g$, which are then compared to the corresponding simulation results.
Since we are dealing with dark-matter only cases, $\Mb$ is set to be infinitesimally small.
We find that the model predictions agree well with the simulation results across the cross-section range.  

While this semi-analytic procedure is already reasonably fast ($\la0.1$ second per system using our publicly available python implementation), it still requires numerical root-finding for determining $\rho_0$ and $\sigma_0$. 
To accommodate semi-analytic frameworks designed for large ensembles of haloes and subhaloes \citep[e.g.,][]{Benson12, Jiang21}, an even faster formula would be useful. 
We find that a \coreNFW profile \citep{Read16} with the scale radius being a fixed fraction of $r_1$ provides decent approximations. 
The \coreNFW profile has an enclosed mass profile given by
\be\label{eq:coreNFW}
M(r) = M_{\rm NFW}(r) \tanh\left(\frac{r}{\rc}\right),
\ee
where $M_{\rm NFW}(r)$ is the enclosed mass of the target NFW profile, and $\rc$ is a characteristic core size. 
We find by trial and error that \coreNFW profiles with $\rc = 0.45 r_1$ fit accurately the SIDM haloes derived from the same target CDM halo across 2 dex in cross section, as shown by the thin dashed lines in \fig{SIDMprofilesPippin}. 
We have verified that this universal approximation holds as long as the system is not in the core-collapse regime, and thus applies to most SIDM haloes with $c\la 20$, $\tage\la14\, \rm Gyr$, and $\sigma_m \la10\, \rm cm^2/g$. 
It breaks down when the baryonic component is not negligible, or when the halo starts to core-collapse, for which a more complicated profile shape is needed. 

%---------------------------------------
\subsection{Comparison with hydro simulations}\label{sec:FIRE}

\begin{figure*}
	\includegraphics[width=0.45\textwidth]{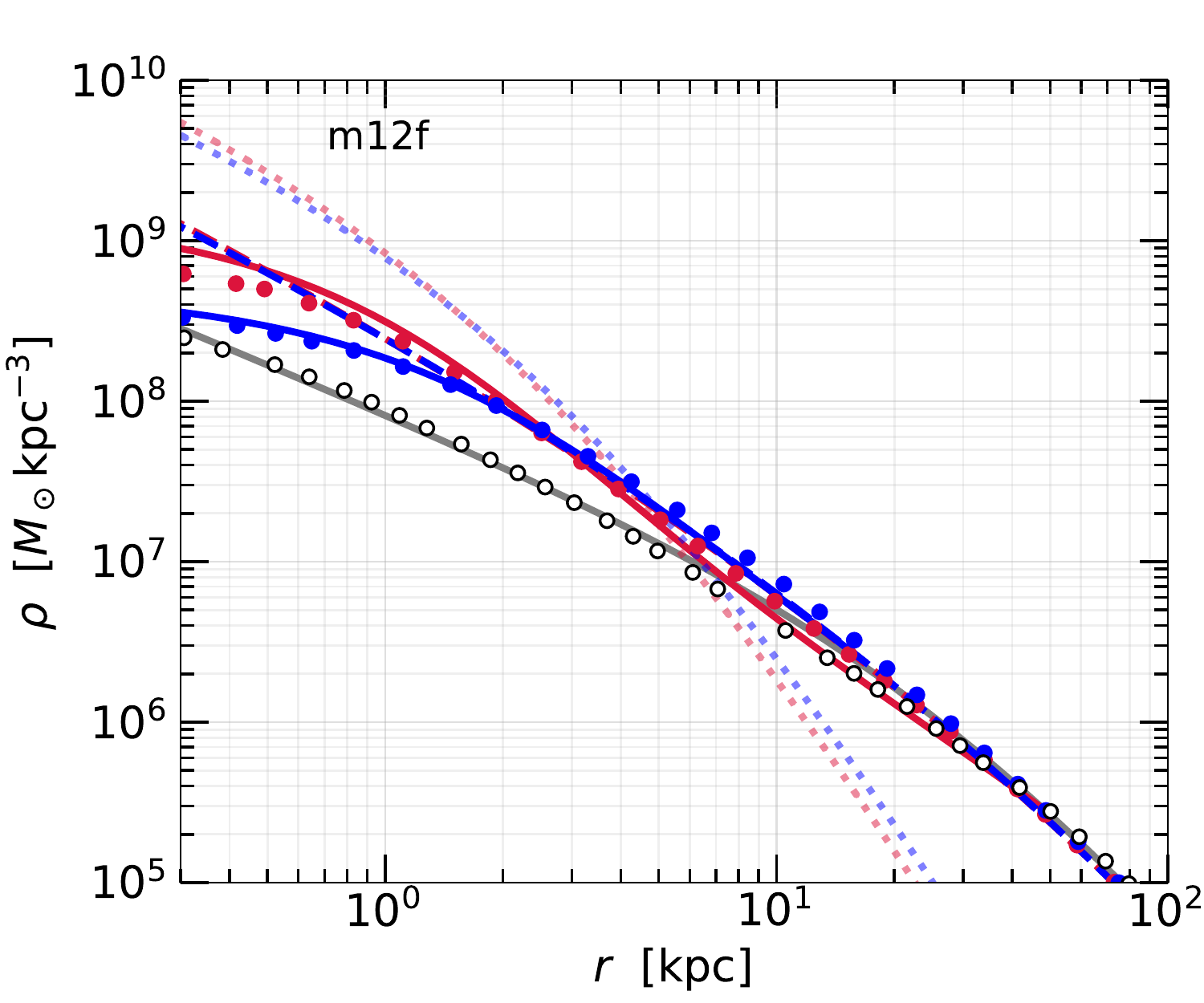}
	\includegraphics[width=0.45\textwidth]{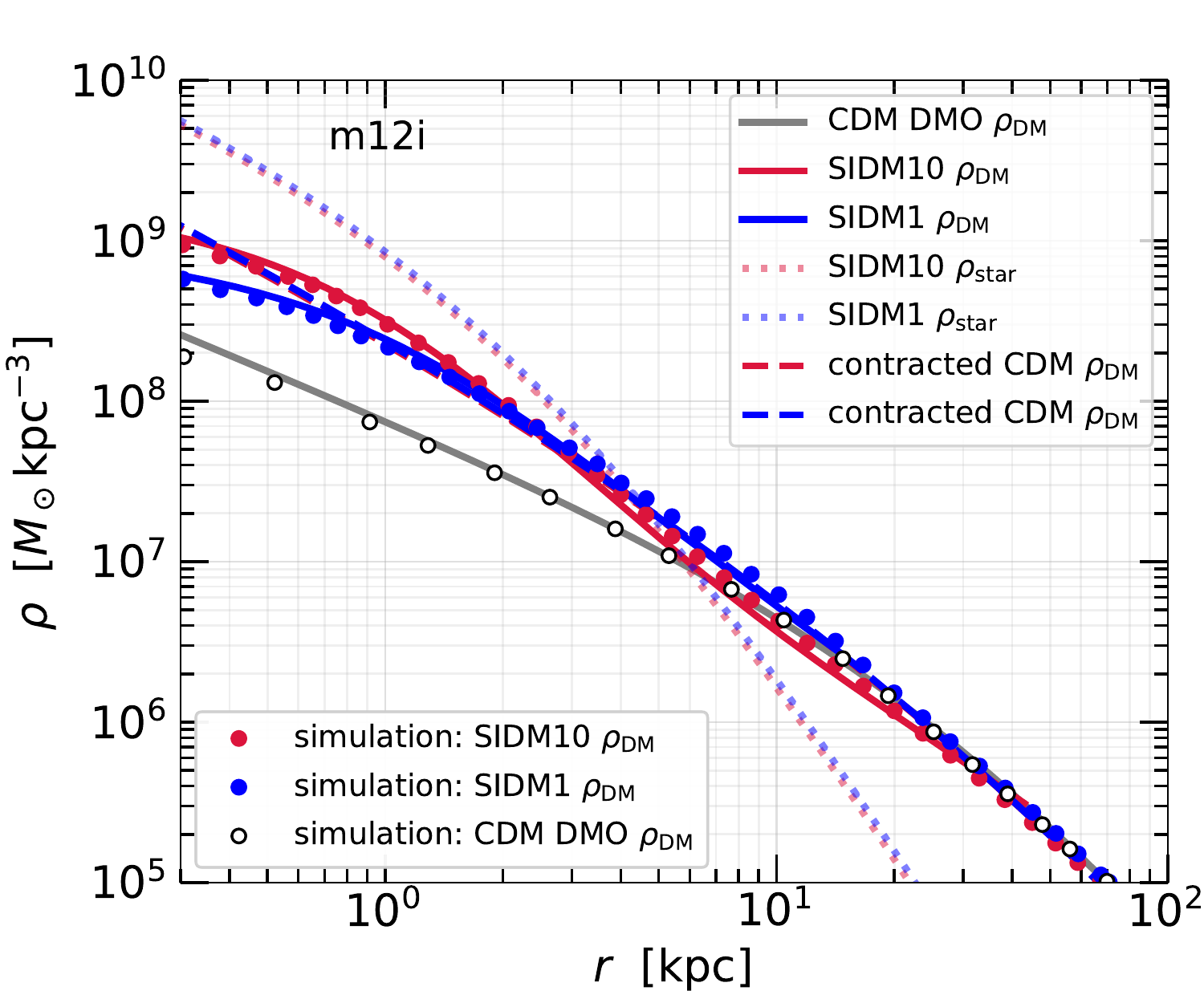}
	\includegraphics[width=0.45\textwidth]{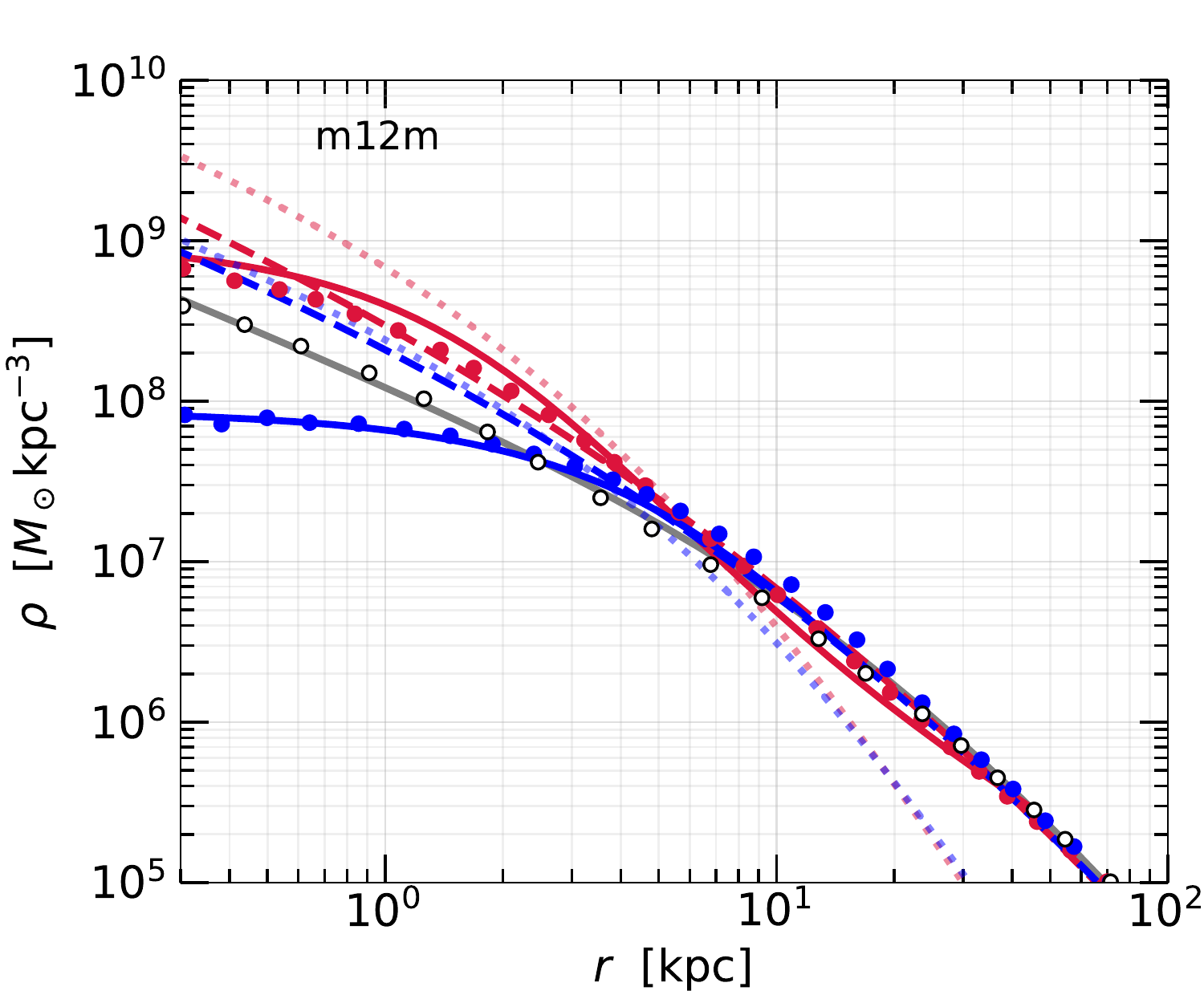}
    \caption{ Comparison of the dark-matter density profiles from the model and from the FIRE2-SIDM hydro-cosmological simulations -- showing three examples of Milky-Way-mass systems at $z=0$: m12f, m12i, and m12m, as in \citet{Sameie21}. 
    The open circles represent the density profiles in the reference CDM dark-matter-only (DMO) runs. 
    The solid grey lines show the best-fit NFW profiles, which are used as inputs in the isothermal Jeans model. 
    The dotted lines represent the best-fit Hernquist profiles of the stellar density distributions in the hydro-simulations. 
    The red and blue colours differentiate the SIDM results for $\sigmam = 10$ (SIDM10) and $1\, \cm^2/\g$ (SIDM1), respectively. 
    The stellar profiles are used as inputs to the model for computing halo contraction. 
    The dashed lines of corresponding colours represent the profiles of the contracted CDM haloes. 
    The filled circles and solid lines of the matching colour represent the profiles from the SIDM simulations and the  corresponding models.
    Overall, the model preditions are in decent agreement with the simulations -- for the SIDM1 run, the central densities at $r\sim1\kpc$ agree at percent level; for SIDM10, the shapes of the simulated profiles are correctly reproduced. }
    \label{fig:SIDMprofilesFIRE}
\end{figure*}

We also compare the model predictions to cosmological hydro simulations, to test its performance when the system is baryon dominated in the centre. 
We use three Milky-Way-mass systems in the FIRE-2 SIDM suite \citep{Sameie21}: m12i, m12f, and m12m, which have virial masses of $M_{200 \rm m}=10^{11.95}$, $10^{12.15}$, and $10^{12.08}\Msun$, respectively, at $z=0$. 
These galaxies are simulated with cross sections of $\sigma_m=1$ and $10\cm^2\g$, and they all have CDM-only reference runs with matched initial conditions which we can use for the model inputs.  
Among the three systems, m12i and m12f have Milky-Way-like sizes of $\rhalf\approx4\kpc$ and a stellar mass of $\Mb\sim10^{10.7}\Msun$, while m12m has a slightly higher stellar mass of $\Mb\sim10^{10.9}\Msun$ and a much more extended stellar distribution of $\rhalf\approx8\kpc$. 
Table 1 of \citet{Sameie21} provides more detailed information of these simulations.  

Again, following the workflow in \se{workflow}, we fit NFW profiles to the CDM-only simulations at $z=0$ and treat the best-fit profiles as the target haloes, as shown by the grey lines in \fig{SIDMprofilesFIRE}.
Then we fit Hernquist profiles to their stellar distributions, as represented by the coloured dotted lines in \fig{SIDMprofilesFIRE}, and use them to model the adiabatic contraction of these haloes. 
We assume these systems formed $\tage=7$ Gyr ago, which is the average formation time of haloes of Milky-Way mass scale.
The predicted SIDM profiles, as shown by the coloured solid lines in \fig{SIDMprofilesFIRE}, match the simulation results fairly accurately. 
For the SIDM1 runs, the central densities are matched at percent levels.
For the SIDM10 runs, while the model slightly overpredicts the central densities, it still correctly captures the shape of the simulated density profiles: there is a relatively flat central core at $r\la1\kpc$, a steep decrease at $r\sim5\kpc$, and a flatter part again at $r\sim r_1\sim 40\kpc$.

The good agreement between the model and the simulations provides insights into the galaxy-halo connection in the context of SIDM. 
In CDM, there are two {\it equally important} competing baryonic effects on halo structure -- on the one hand, the galactic potential makes the halo contract and become more cuspy; on the other hand, supernovae-driven outflows heat the potential well and flatten the central density. 
The net effect of the competing mechanisms depend sensitively on details of the subgrid physics for star formation and supernovae \citep[e.g.,][]{Bose19}. 
The SIDM simulations here also include both of the competing mechanisms, but the model only considers halo contraction and ignores stellar feedback. 
Hence, the fact that good agreement is still achieved between the model and the FIRE2-SIDM simulations implies that the core-formation effect from supernovae is subdominant and overwhelmed by the effect of the SIDM halo in the presence of the baryonic potential (see also \citeauthor{Sameie21} for discussion). 
It is therefore reasonable to speculate that SIDM simulations are not sensitive to the sub-grid baryonic physics for certain ranges of SIDM parameters.
This should be better tested with hydro+SIDM simulations with varied strength of feedback. 

%---------------------------------------------------------------------------------------------------------
\section{SIDM halo response}\label{sec:HaloResponse}

In this section, we use the model for quantitative analysis of the SIDM halo response. 
We express the halo structures as functions of the baryonic mass fraction ($\Mb/\Mv$) and the baryonic compactness ($\rhalf/\Rv$), and also take this opportunity to show the importance of considering adiabatic halo contraction. 

%---------------------------------------
\subsection{Enhanced structural diversity in SIDM}\label{sec:response}

\begin{figure*}	
	\includegraphics[width=0.7\textwidth]{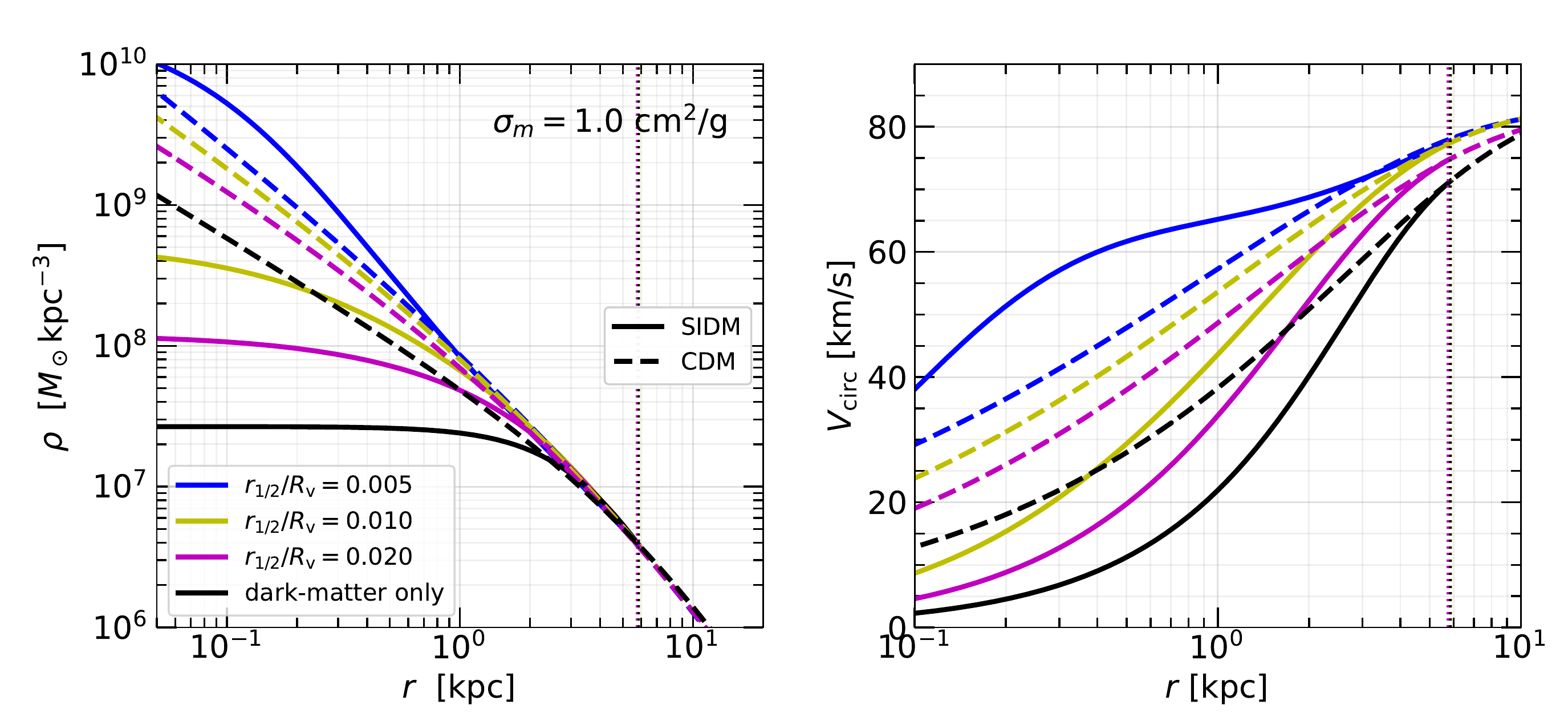}
	\includegraphics[width=0.7\textwidth]{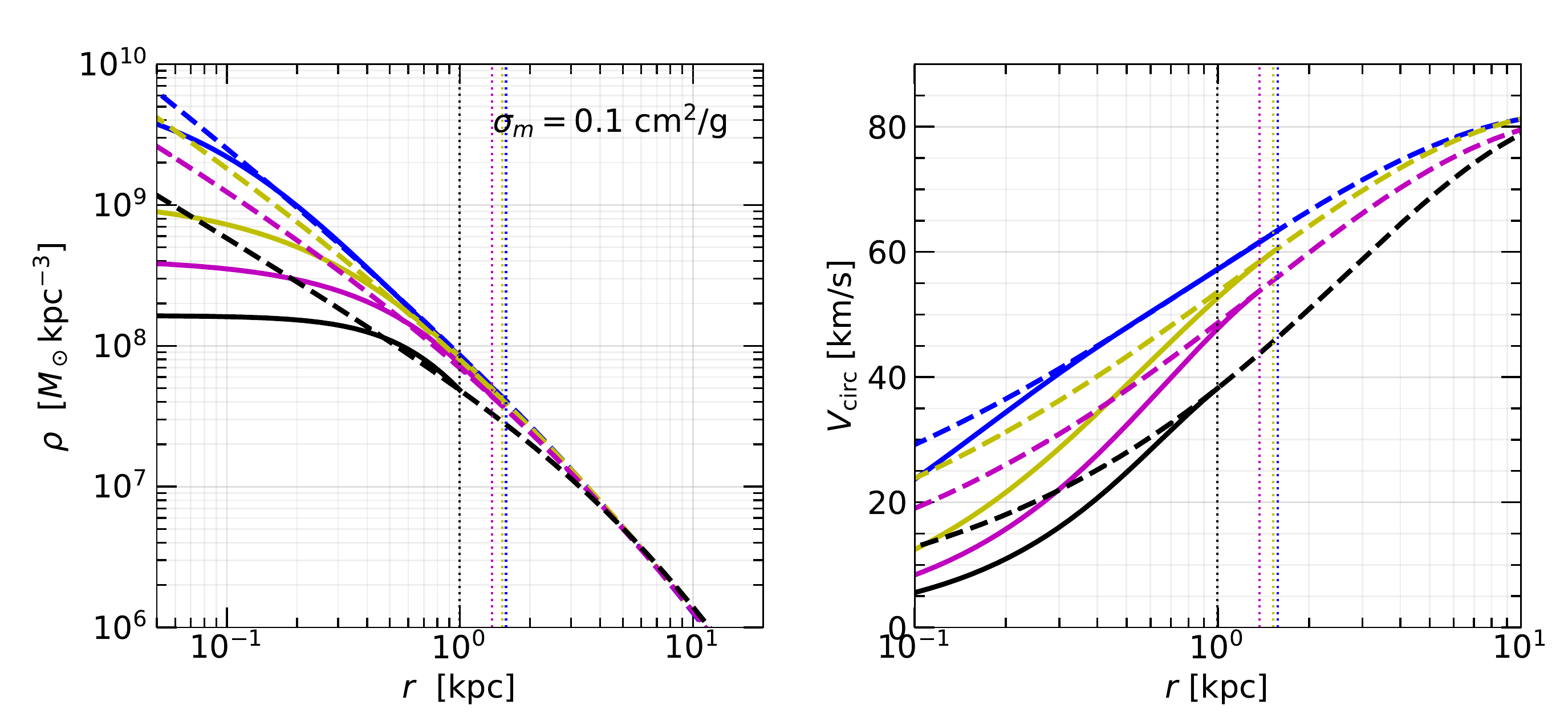}
    \caption{ Illustration of the high sensitivity of SIDM halo response to baryonic potentials.
    The left-hand panels and right-hand panels show the density profiles and circular velocity profiles, respectively, of SIDM haloes (solid lines) and CDM counterparts (dashed lines). 
    For all the cases, we keep fixed the virial mass of $\Mv=10^{11}\Msun$, the target concentration of $c=10$, and the galaxy mass of $\Mb=10^{9}\Msun$, only varying the galaxy size. 
    The colours differentiate the half-mass radii of $\rhalf=0.005$, 0.01, and 0.02$\Rv$, as indicated, or equivalently, $\rhalf\approx0.47$, 0.95, and 1.9 kpc -- these are representative of compact, normal, and ultra-diffuse dwarf galaxies. 
    The upper and lower panels show the results for cross sections of $\sigmam=1\cm^2/\g$ and $0.1\cm^2/\g$, respectively.
    The vertical dotted lines mark the positions of $r_1$ for the corresponding colour.
    Clearly, SIDM haloes are more sensitive to baryonic compactness than their CDM counterparts. 
    The strong difference in the inner halo is not driven by the difference in $r_1$, which is actually negligible for $\sigmam=1\cm^2/\g$ or larger; instead, it arises from the difference in $\Vc(r)$, or more precisely in the derivatives of the gravitational potential $\rmd\Phi/\rmd r = \Vc(r)^2/r$ and $\rmd^2\Phi/\rmd r^2$, as hinted from the right-hand panels.
    }
    \label{fig:HaloResponseSensitivity}
\end{figure*}

Zoom-in hydro-simulations have hinted that SIDM haloes are more responsive to the presence of a baryonic distribution (rather than baryonic feedback) than their CDM counterparts. 
Here, we use the isothermal Jeans model to show this more explicitly.

First, we vary the size of the baryonic component while keeping the total mass and baryon mass fixed at $\Mv=10^{11}\Msun$ and $\Mb=10^{9}\Msun$ -- these values are typical of bright dwarf galaxies such as the LMC or sub-$L^\star$ galaxies which exhibit the most dramatic structural diversity.
We also keep the halo age and the target-halo's concentration fixed at typical values of $\tage = 10$ Gyr and $c=10$. 
We run the model for two cross sections, $\sigmam=1$ and $0.1\cm^2/\g$.
We perform control-experiments to get the CDM references, i.e., starting from the same target halo and the same galaxy as used for the SIDM calculations, and simply compute the adiabatically contracted CDM halo profiles. 
\fig{HaloResponseSensitivity} shows the comparison. 
The sensitivity of the halo response in the SIDM models is indeed much higher than that of the reference CDM cases. 
Notably, the inner SIDM density slope (evaluated at, e.g., $r\sim0.5\kpc \approx 0.5\%\Rv$) can be flat, equally cuspy, or cuspier than that of the reference CDM profile, depending on whether the galaxy is diffuse ($\rhalf=2\kpc$), normal ($1\kpc$), or compact ($\rhalf\simeq0.5\kpc$). 
The range of the central densities, e.g., evaluated at $r=0.1\kpc$, of the CDM results is only 0.5 dex, while that of the SIDM models spans more than an order of magnitude. 
%This point has been made in the papers discussing 

This remarkable diversity in halo response is not driven by the difference in the characteristic radius $r_1$.
In fact, for $\sigmam\ga1\cm^2/\g$, the $r_1$ values are similar across the different galaxy sizes, as shown by the vertical dotted lines in \fig{HaloResponseSensitivity}. 
Only for cross sections as small as $\sigmam\sim0.1\cm^2/\g$, $r_1$ becomes comparable to the galaxy size and differs significantly depending on the latter.
Even here, $r_1$ occurs where the halo density profiles converge, so the dramatic difference in the inner halo cannot be attributed to that of $r_1$ or of the local density $\rho(r_1)$. 
The structural diversity must then arise from the difference in the enclosed mass profile, or $\Vc(r)$, as shown in the right-hand panels of \fig{HaloResponseSensitivity}. 
A small change in the baryonic size results in amplified differences in the gradient and the Laplacian of the potential, $\rmd\Phi/\rmd r = \Vc(r)^2/r$ and $\rmd^2\Phi/\rmd r^2$, which are leading terms in the Jeans-Poisson equation (\eq{JeansPoissonSimplified}) underlying the whole model.
%This highlights the importance of taking into account halo contraction in the model, because otherwise $\Vc(r_1)$ will not be different. 

The structural diversity of bright dwarf galaxies ($\Mb\approx10^{8-9}\Msun$) has drawn a lot of attention recently. 
Notably, these galaxies span two orders of magnitude in size and exhibit a wide range of morphologies, including  compact dwarfs with $\rhalf$ as small as $\sim0.1\kpc$ and ultra-diffuse galaxies with $\rhalf$ up to $10\kpc$. 
The structural diversity is also manifested in the logarithmic density slope $s\equiv\rmd\ln\rho/\rmd\ln r$  near the centre ($r \la 1\kpc$), as inferred from baryonic kinematics. 
For example, as \citet{Relatores19} summarized, $s$ ranges between $0$ and $1.5$ for galaxies with $\Mb\sim10^9\Msun$. 
It is challenging for hydro+CDM models to fully explain such a dramatic extent of structural diversity, especially given that both the galaxy size and the inner halo structure exhibit wide ranges. 
Recently, \citet{Zentner22} demonstrated that SIDM and feedback-affected CDM models are equally better than a CDM model in explaining the halo structural diversity as seen in the SPARC survey \citep{Lelli16}, however, the prevalence of compact bright dwarfs with $\rhalf\la1\kpc$ remains a challenge for hydro-CDM simulations featuring strong feedback \citep[e.g.,][]{Jiang19}.
Here, galaxy size is an input of the model, so we do not provide an explanation for the size diversity, but {\it we have clearly shown that SIDM models have the virtue of making the two aspects strongly coupled, such that if there is an explanation for the size diversity, it explains automatically the range of DM density slopes. }

\begin{figure*}	
\includegraphics[width=\textwidth]{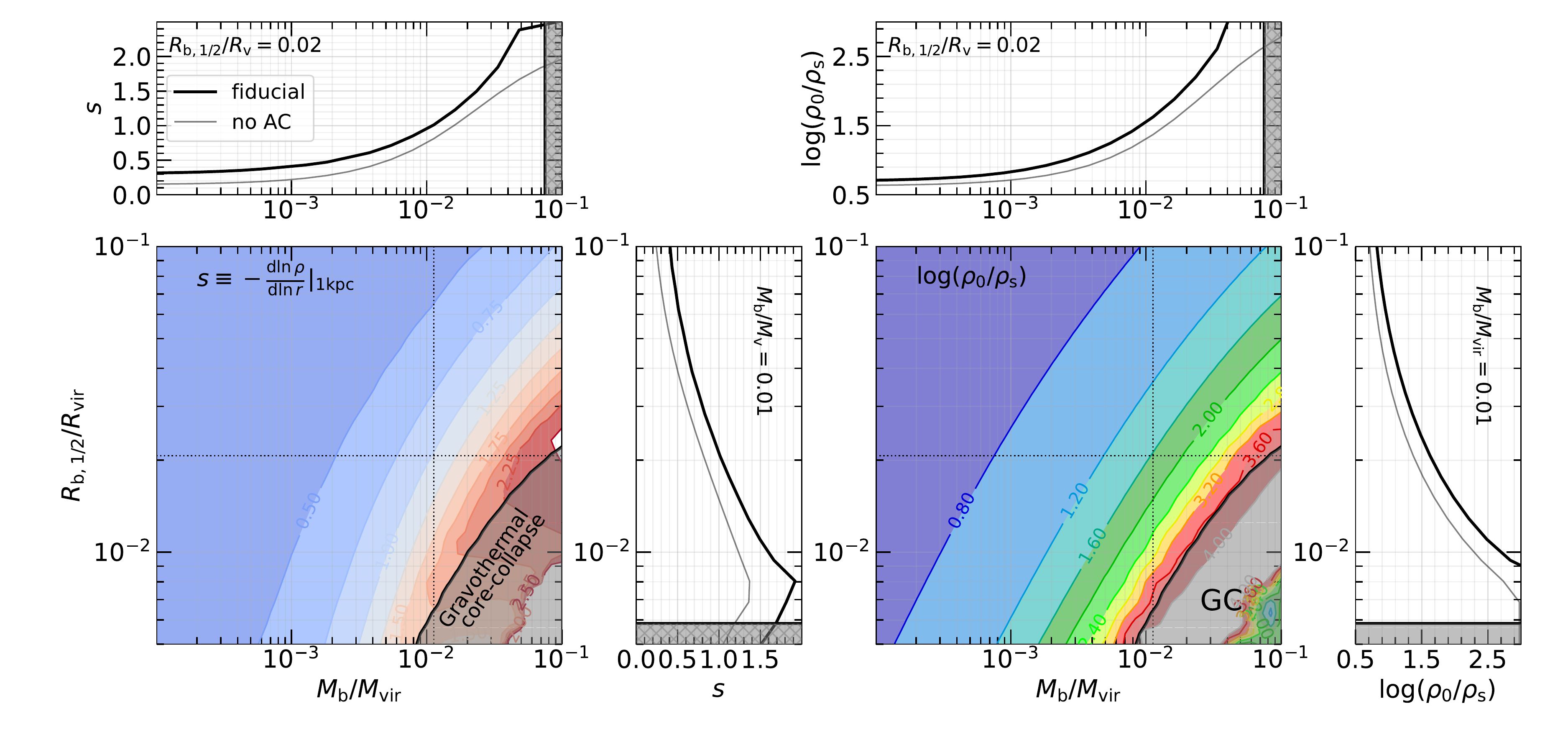}
    \caption{ The inner logarithmic density slope (left) and central density (right) of SIDM haloes as functions of the baryonic mass fraction $\Mb/\Mv$ and the galaxy size in units of the virial radius $\rhalf/\Rv$.
    Here we choose $\sigmam=1\cm^2/\g$ and adopt a target halo formed $\tage=10\Gyr$ ago with present-day virial mass $\Mv=10^{11}\Msun$ and concentration $c=10$. 
    The slope $s\equiv - \rmd\ln\rho/\rmd\ln r$ is evaluated at $r=1\kpc$, and the central density is expressed in units of the NFW scale density $\rhos$.
    The main panels are the contour maps of $s$ and $\log(\rho_0/\rho_s)$, with the contour-level values indicated. 
    The thick black lines in the top and side panels show one-dimensional slices of the main panels, with one of the baryonic properties fixed at the values indicated by the dotted lines in the main panel. 
    The thin grey lines in the top and side panels show the results without considering adiabatic halo contraction -- these are equivalent to the `inside-out' models of \citet{Robertson21}.
    Focusing on our fiducial model with adiabatic contraction, the density slope increases from $\sim0.3$ to $2$ as $\Mb/\Mv$ increases from $10^{-4}$ to 0.04 or as $\rhalf/\Rv$ decreases from $0.1$ to $0.005$, for the specific slices.
    For $\Mb/\Mv\sim 0.01$ and $\rhalf/\Rv\sim 0.02$, representative of Milky-Way-mass galaxies according to abundance matching, the SIDM density profile is actually rather similar to the CDM case with an inner slope of $\sim1$.
    Accounting for adiabatic contraction makes the central density up to four times higher (e.g., at $\Mb/\Mv=0.04$) and the central density slope $\sim$30\% steeper.
    In the lower right corner of the space, an isothermal solution can no longer be achieved, manifesting the speed-up of gravothermal core-collapse (GC). 
     }
    \label{fig:HaloResponse}
\end{figure*}

Second, we extend the above exercise by scanning a wide range in the space spanned by the baryonic mass fraction and galaxy compactness, and thus more systematically describe the SIDM halo response. 
Still adopting $\sigmam=1\cm^2/\g$ and a target CDM halo of $\tage=10\Gyr$, $\Mv=10^{11}\Msun$, and $c=10$, we vary $\Mb/\Mv$ from $10^{-4}$ to 0.1, and $\rhalf/\Rv$ from $0.004$ to 0.1. 
We express the halo structure in terms of the inner density slope $s\equiv - \rmd\ln\rho/\rmd\ln r$  evaluated at $r=1\kpc$, and the central density $\rho_0$ in units of the NFW scale density $\rhos$.
The results are shown in \fig{HaloResponse}.
The main panels of \fig{HaloResponse} show the contour maps of $s$ and $\rho_0/\rhos$ in the 2D baryon-property space.
The top panels and side panels show the 1D slices of the 2D map with either of the baryonic quantities fixed (at $\Mb/\Mv=0.01$ or $\rhalf/\Rv=0.02$).
Clearly, the SIDM halo becomes more dense and cuspy as the galaxy becomes more massive and compact.  
Although we have used a massive dwarf halo for illustration, the result applies to other mass scales as well since we have expressed the baryonic properties in units of the virial quantities.  

Hydro-cosmological zoom-in simulations have shown that, for Milky-Way-like systems, SIDM halo profiles are rather similar to their CDM counterparts down to quite small radii.
This can be seen for example in m12f and m12i in \fig{SIDMprofilesFIRE}, and it has motivated some semi-analytic studies to assume NFW profiles for their Milky-Way sized SIDM host halo when studying the satellite galaxies \citep[e.g.,][]{Jiang22}. 
Here we can easily check the validity of this assumption in \fig{HaloResponse}.
Abundance-matching studies have shown that a Milky-Way-mass system typically has a stellar-to-total-mass ratio of $\sim1$ per cent \citep[e.g.,][]{Moster13}, and a half mass radius that is $\sim2$ per cent of the host-halo virial radius \citep[e.g.,][]{Somerville18}.
For these representative values, as can be seen in \fig{HaloResponse}, the SIDM profile indeed has an inner logarithmic density slope very close to the NFW value of $s \sim 1$. 

%---------------------------------------
\subsection{Necessity of considering adiabatic contraction}\label{sec:NeedForAC}

\citet{Robertson21} also studied the isothermal Jeans model in detail and made comparisons with cosmological simulations.
There, the authors adopted an {\it inside-out} fitting scheme.
That is, different from what we do here, they start from an isothermal core profile defined by $\rho_0$ and $v_0$ in the centre, evaluate $r_1$ using the core profile, and find the NFW profile on the outskirt that smoothly joins the core at $r_1$.
In this regard, our workflow as described in \se{workflow} is called the {\it outside-in} approach \citep[e.g.,][]{Sagunski21}. 
As \citet{Robertson21} noted, in the inside-out approach, the outer halo is completely determined by the NFW profile and there is no freedom to incorporate contraction.
That said, it is still able to capture the effect of baryonic potential on the SIDM profile partially, via the baryonic terms in the Jeans-Poisson equation, \eq{JeansPoisson}.
It is just not entirely self-consistent, as the baryonic potential will affect the entire halo, making the outer part also deviate from NFW. 

Here, with the outside-in approach, we can quantify the difference made by including adiabatic halo contraction. 
We emulate the inside-out model by skipping the halo contraction step of our workflow and only consider the baryonic potential in the Jeans-Poisson equation. 
The difference is shown in the top and side panels of \fig{HaloResponse} -- the thick black lines show the halo response from the fiducial model, and the thin grey lines show the result skipping halo contraction (with everything else the same). 
As can be seen, accounting for adiabatic contraction does not introduce a big difference for galaxies of $\Mb/\Mv\la 0.001$ or for diffuse systems of $\rhalf/\Rv\ga0.04$; however, for massive and compact systems, the central density in our fiducial model can be up to four times higher (see e.g., the result at $\Mb/\Mv=0.04$), and the central density slope can also be different by up to 30\%.  
%AC affects the results in two aspects, first, it slightly changes the profile and, depending on the cross section and halo age, affect the $r_1$radius slightly. But more importantly, it increases the enclosed mass $\Vc(r_1)$ significantly. 
In short, for massive and compact systems, an explicit adiabatic-contraction treatment must be included for accurate results; for diffuse and dark-matter dominated systems, considering the baryon potential in the Jeans Poisson equation provides results that are close enough.
%We think that the decision should not be naively based on whether the system has an $r_1$ comparable to the baryonic size, because even in this case, the contracted halo profile can happen to have a local density similar the NFW halo.

%---------------------------------------------------------------------------------------------------------
\section{Discussion}
\label{sec:discussion}

In this section, we first compare the isothermal Jeans model to the more sophisticated gravothermal fluid model, which also predicts SIDM halo profiles and is studied extensively in the literature. 
Then, we study the facilitation of gravothermal core-collapse by the inhabitant galactic potential, and use the isothermal Jeans model to predict the regime of core-collapse in the space of galaxy mass fraction versus galaxy compactness. 

%---------------------------------------
\subsection{Comparison with gravothermal fluid evolution}\label{sec:gravothermal}

The isothermal Jeans model assumes a system to be in approximate equilibrium, whereas with dark-matter self-interactions, the system is never in strict equilibrium. The full hydrodynamical evolution can be described by the gravothermal fluid model, which is extensively studied in a series of seminal works \citep{LE80, Balberg02, Koda11, Pollack15, Essig19, Nishikawa20}.
This method treats SIDM as a gravothermal fluid, and solves a set of coupled partial differential equations for the evolution of the spherically symmetric profiles of mass $M(r,t)$, density $\rho(r,t)$, velocity dispersion $v(r,t)$, and the luminosity of the radiated heat $L(r,t)$ -- 
\bad\label{eq:fluid}
\frac{\partial \Mt}{\partial \rt} &= \rt^2 \rhot, \\
\frac{\partial (\rhot\vt^2)}{\partial \rt} &= -\frac{\Mt\rhot}{\rt^2}, \\
\frac{\partial \Lt}{\partial \rt} & = -\rt^2 \rhot \vt^2 \left( \frac{\partial}{\partial \ttilde} \right)_{\Mt} \ln\left(\frac{\vt^3}{\rhot}\right), \\
\Lt & = -\frac{3}{2} \rt^2 \vt \left( \frac{a}{b}\sigmatm^2 + \frac{1}{C\rhot\vt^2} \right)^{-1} \frac{\partial \vt^2}{\partial \rt}.\\
\ead
These equations describe mass conservation, hydrostatic equilibrium, the first law of thermodynamics, and heat conduction, respectively, where $a=4/\sqrt{\pi}$, $b=25\sqrt{\pi}/32$, and $C$ is a calibration parameter of order unity. 
Following \citet{Koda11} and \citet{Nishikawa20}, we have expressed the equations with the dimensionless quantities:  
$\rt\equiv r/\rs$, $\rhot\equiv \rho/\rhos$, $\Mt\equiv M/M_0$ with the mass scale $M_0=4\pi\rs^3\rhos$, $\sigmatm\equiv \sigmam/\sigma_{m0}$ with the cross-section scale $\sigma_{m0}=1/\rs\rhos$, $\vt\equiv v/v_0$ with the velocity scale $v_0 = \sqrt{GM_0/\rs}$, $\Lt\equiv L/L_0$ with the luminosity scale $L_0\equiv GM_0^2/\rs t_0$, and $\ttilde\equiv t/t_0$ with the time scale $t_0 = 1/a\sigmam v_0 \rhos$.
This assumes that the initial profile $\rho(r,t=0)$ is NFW, with scale radius $\rs$ and scale density $\rhos$.
With the dimensionless quantities, we have the convenience that the density-profile evolution is self-similar as long as we are in the long-mean-free-path\footnote{That is, when the mean free path of scattering, $\lambda=1/\rho v$, is larger than the gravitational scale length, $H=\sqrt{v^2/4\pi G \rho}$} regime, and thus the result is almost independent of the cross section or the initial NFW concentration when expressed in $\rhot(\rt,\ttilde)$.
This is illustrated in \citet{Balberg02} and in Appendix C of \citet{Nishikawa20}.

There are a few differences between the isothermal Jeans model and the fluid model.
First and foremost, conceptually, the fluid model gives the full (time-dependent) solution to the Boltzmann Equations with an assumed conductivity; while the isothermal model approximates the instantaneous profile as being in equilibrium, and therefore does not have time evolution per se other than a dependence on halo age. 
Second, the isothermal model is only applicable to the isothermal-coring stage and the onset of gravothermal core-collapse; while the fluid model can follow the evolution well into core-collapse. 
Third, solving the fluid equations requires discretizing the spherical halo and is relatively computationally expensive; whereas the isothermal model only requires performing the minimization at $r_1$, and within each iteration, the numerical integration of the Jeans-Poisson equation is quite fast. 
The speed advantage makes it easier for incorporating into large semi-analytic frameworks. 
Fourth, the fluid model only considers the dark-matter component, at least as presented in the literature so far; while the isothermal model easily accounts for baryonic effects by including baryonic terms in the Jeans-Poisson equation and by considering adiabatic contraction. 
For this reason, when we compare the the two models, we focus on the dark-matter-only setups.\footnote{It is in principle possible to include a static baryon component in the second equation of \eq{fluid} and thus make the fluid model capture baryon effects as well, but this is beyond the scope of this work.}
Finally, the fluid model can easily adapt to velocity-dependent cross sections -- one can simply plug a $v$-dependent cross section $\sigmam(v)$ in the fourth equation of \eq{fluid}; while the isothermal model evaluates the $r_1$ radius using the instantaneous cross section, and thus ignores any $v$-dependence. 
For typical particle-physics models, the $v$-dependence effectively makes the cross section larger in the past and thus makes the isothermal coring faster \citep{Nadler20}. 
That said, if we know the growth history of the target CDM halo including the velocity dispersion profile as a function of redshift $v(r,z)$, then we can solve for an $r_1(\tage)$ that includes the time dependence:
\be\label{eq:vDependentR1}
1=\int_{0}^{\tage}\rho(r_1,t)v(r_1,t)\sigmam[v(r_1,t)]\rmd t,
\ee
where $t(z)$ is the lookback time.
We can therefore perform the isothermal Jeans modeling for each time and construct a density-profile evolution $\rho(r,z)$ that approximates the case of a $v$-dependent cross section. 

\begin{table}
\caption{Comparison of the two 1D models of self-interacting DM haloes -- the isothermal Jeans model versus the gravothermal fluid model. See \se{gravothermal} for details}
 \begin{threeparttable}
\begin{tabular}{lll} 
\hline
\hline
  & Isothermal & Gravothermal \\
\hline
 & {\it similarities} &\\
Operation target  &  CDM halo   &  CDM halo \\
Applicable before core-collapse & yes & yes\\
\hline
 & {\it differences} &\\
Speed   & fast  &  slow \\
Applicable after core-collapse & no & yes\\
Captures baryonic effect & yes$^{\rm a}$ & no$^{\rm b}$\\
Support $v$-dependent $\sigmam$ & no$^{\rm c}$ & yes \\ 
\hline
\end{tabular}
\begin{tablenotes}
\item[a] It captures the gravitational effect of the baryonic potential, not the baryonic feedback. 
\item[b] In principle, one can add a static baryonic term in the second equation of \eq{fluid}, such that the gravothermal fluid model can also capture the halo response to the baryonic component.
\item[c] However, velocity dependence effectively makes the cross section larger in the past, so if given the growth history of the target CDM halo, we can redefine $r_1$ with \eq{vDependentR1} and perform the isothermal Jeans modeling for each time.
\end{tablenotes}
\end{threeparttable}
\label{tab:IsothermalVersusGravothermal}
\end{table}

\begin{figure}
	\includegraphics[width=0.49\textwidth]{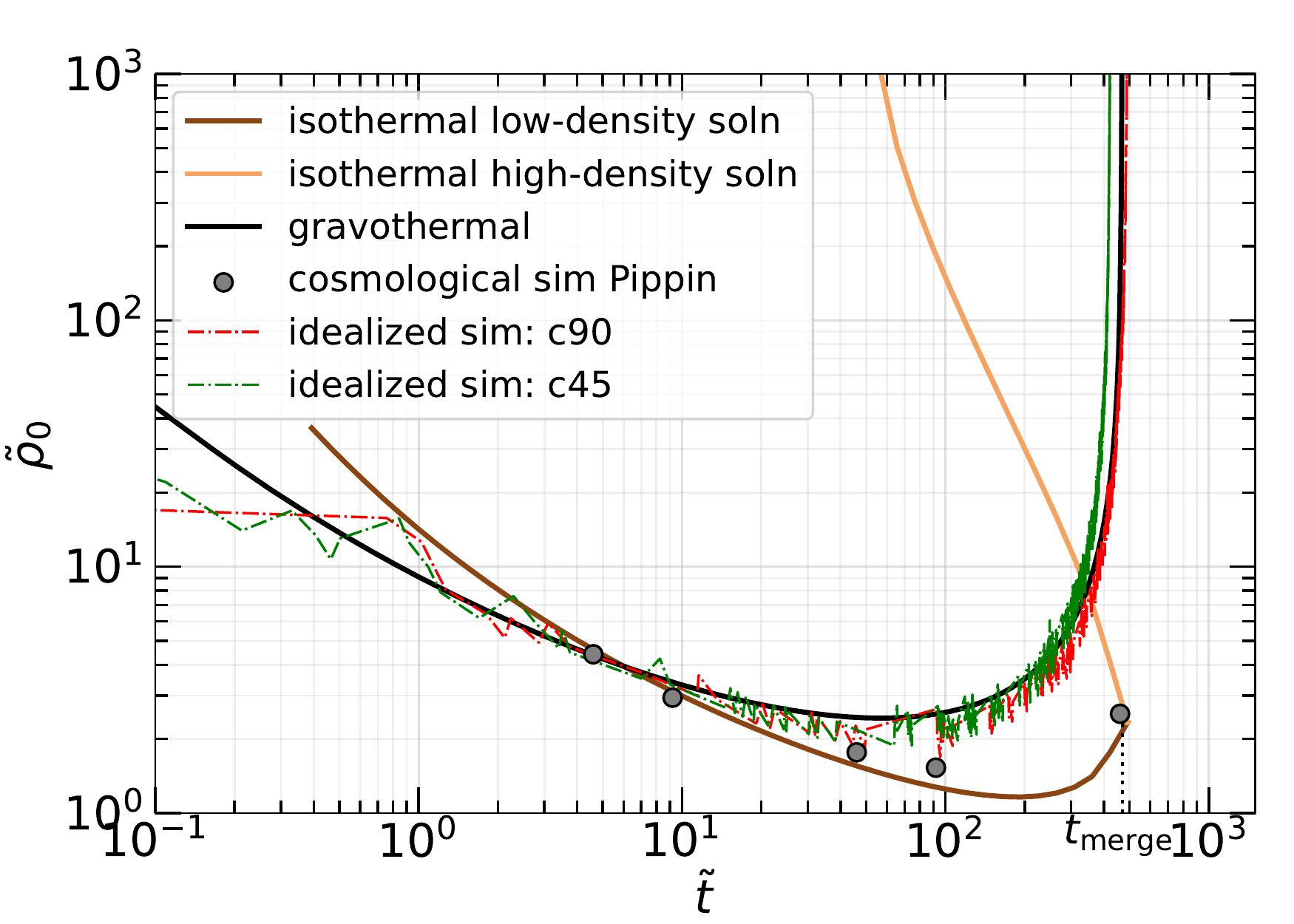}
    \caption{ Comparison of the gravothermal fluid model (black solid line) and the isothermal Jeans model (brown solid line) in terms of the (dimensionless) central density $\rhot_0\equiv\rho_0/\rhos$ as a function of time $\ttilde\equiv t/t_0$. See \se{gravothermal} for the definitions and the details of the calculations. 
    Note that the details of the target CDM halo or the cross section have little impact on the dimensionless $\rhot_0(\ttilde)$ track. 
    Simulation results are overplotted for comparison -- the grey circles represent the Pippin {\it cosmological} simulations of different cross sections $\sigmam=0.5$, 1, 5, 10, and 50 $\cm^2/\g$ at $z=0$; the red and green dash-dotted lines represent the {\it idealized} isolated simulations starting from NFW profiles with $c=45$ and $90$, and with $\sigmam=10\cm^2/\g$.   
    The isothermal model agrees better with the cosmological results, while the fluid model agrees with the idealized simulations -- their difference likely originates from whether the target CDM halo is used as an {\it initial condition} or as a {\it boundary condition} (see \se{gravothermal} for discussion). 
    The orange solid line represents the usually-discarded high-density solution of the isothermal model. 
    The point when the high-density and low-density solutions merge coincides with when gravothermal core-collapse speeds up and the core temperature is well above the velocity dispersion of the CDM-like outskirt (see \se{GC}). 
    %The orange dashed line represents the mirroring of the high-density track with respect to the merging time ($t_{\rm merge}$) -- it seems to phenomenologically provide a smooth extrapolation of the isothermal solution into the core-collapsed regime where the model is in principle not applicable. 
    %Core-collapse completes at $\tilde{t}_{\rm GC}= 2\tilde{t}_{\rm merge}$.
    }
    \label{fig:IsothermalVersusGravothermal}
\end{figure}

We summarize these similarities and differences of the two methods in \tab{IsothermalVersusGravothermal}. 
Overall, the isothermal Jeans model is simplistic yet much faster. 
In \fig{IsothermalVersusGravothermal}, we compare the two models in the space of the dimensionless central density $\rhot_0$ versus the dimensionless time $\ttilde$. 
For the fluid model, $\rhot_0 (\ttilde)$ is simply obtained by solving \eq{fluid}.
We have followed the numerical method as detailed in \citet{Nishikawa20}, starting from the NFW profile of the $z=0$ CDM Pippin halo (i.e., solid grey line in \fig{SIDMprofilesPippin}), adopting a cross section of $\sigmam=5\cm^2/\g$, and using $C=0.75$ as calibrated to idealized simulations \citep{Koda11}.
Despite the specific choices, we emphasize that the cross section, the details of the NFW profile, or the exact value of $C$ as long as it is between 0.5 and 1, has weak impact on the result in this dimensionless space in the core-forming regime. 
For the isothermal model, in order to construct the `time evolution', we repeat the exercise for a series of halo age $\tage$ and plot $\rhot_0$ versus $\ttilde\equiv\tage/t_0$. 
The same target CDM halo and cross section are used for both methods.
Again, these details are largely irrelevant for this dimensionless parameter space due to the self-similar nature of the density evolution in the core-formation regime, and we have verified with the isothermal Jeans model  that it predicts a universal track in the $\rhot_0$-$\ttilde$ space for different $\sigmam$.  
For the isothermal model, in addition to the default, low-density solution, we also record the high-density solution, and display both solutions in \fig{IsothermalVersusGravothermal}. 
We reiterate that only the low-density solution is supposed to be comparable to the simulation results or the fluid model predictions.

As can be seen, both models show a similar qualitative behavior -- an isothermal core grows  as the density keeps decreasing; then the central density reaches a minimum and turns around, manifesting the onset of gravothermal core-collapse. 
However, there is a clear difference: with the isothermal model, the core develops faster, and reaches a minimum central density that is $\sim2$ times lower than that predicted by the fluid model, at a slightly later time.  
This difference cannot be attributed to the calibration parameter $C$. In fact, smaller (larger) $C$ makes the turn-around of $\rhot_0$ occur later (earlier), but it has little impact on the steepness of the isothermal-coring stage. 

What causes the difference? Which model is more accurate? 
To get some clues, we compare the model predictions to simulation results of different kinds. 
First, we compare to the {\it cosmological} Pippin $N$-body simulations of \citet{Elbert15}. 
Following \citet{Essig19}, an `evolutionary' track $\rhot_0$($\ttilde$) can be constructed using the simulation results all at the same time of $z=0$.
This is because the dimensionless time $\ttilde\equiv t(z)/t_0\propto\sigmam^{-1}$, where the cosmic time at $z=0$ is $t=13.7\Gyr$ for the Pippin cosmology, so different cross sections correspond to different dimensionless times. 
Specifically, the Pippin halo was run with cross sections $\sigmam=0.5$, 1, 5, 10, and 50 $\cm^2/\g$, and the central densities at $z=0$ are $\rho_0=7.5$,  5.0, 3.0, 2.6, and 4.3$\times10^{7}\Msun\kpc^{-3}$, respectively. 
The CDM counterpart has $\rhos=1.7\times10^7\Msun\kpc^{-3}$ and $\rs=2.7\kpc$.
Hence, the dimensionless central densities are $\rhot_0 = 4.4$, 2.9, 1.8, 1.5, and 2.5, which are reached at the dimensionless times of $\ttilde= 4.6$, 9.2, 46, 92, and 460, respectively. 
Interestingly, the isothermal Jeans model, albeit simplistic, agrees with the cosmological Pippin simulations very well.
Notably, the steeper isothermal coring at $\ttilde\la100$ is the same, and the last simulation data point at $\ttilde=460$, which exhibits core-collapse, is almost on top of the model prediction.
This time happens to be when the low-density solution and the high-density solution merge, beyond which the isothermal Jeans model is no longer applicable. 
Mathematically, for a continuously evolving quantity (such as the central density $\rhot_0$) that has two solutions, any transition between the solutions must be continuous and therefore any continuous parameter (such as time $\tilde{t}$) must enable a smooth transition between the solutions. 
In this sense, the transition is when the density increases and that is the onset of core collapse. Physically, beyond this time, a negative velocity-dispersion gradient starts to develop so the isothermal assumption breaks (see \se{GC} and \app{TwoSolutions} for more discussion). In practice, the merging point is manifested by where the `stitching error' $\delta^2$ can no longer be minimized to zero.

Second, we compare the model predictions to {\it idealized} SIDM $N$-body simulations of isolated haloes.
To this end, we simulate with the \texttt{Arepo} code \citep{Springel10, Weinberger20} two Milky-Way sized haloes, which are initialized with NFW profiles at $z=0$ with $c=45$ and 90, respectively, and are evolved with a self-interacting cross section of $10\cm^2/\g$. The details of the simulations are provided in \app{idealized}. 
Again, $\rhot_0(\ttilde)$ is not sensitive to the details of the target halo or the cross section, and the high concentrations are chosen simply to facilitate the gravothermal evolution and shorten the computation time. 
Clearly, the fluid model agrees well with the idealized simulations, whereas the isothermal model agrees better with cosmological results. 

We hypothesize that the difference originates from how the target CDM halo is used in the modeling. 
Specifically, in the fluid model, the present-day target CDM halo is used to {\it initialize} the system. 
That is, there are two implicit assumptions here: first, for the entire history of the target halo up until $t=0$, dark matter remains collisionless, and only at $t > 0$, DM becomes self-interacting; second, the halo stops mass accretion and evolves in isolation at $t > 0$. 
As such, the profile we obtain at time $t$ ($t>0$) is virtually that of an isolated system at a future cosmic time of $t$ + the age of the Universe today, with the effect of self-interactions during the entire assembly of the target halo not taken into account.
It is therefore not surprising that the fluid model disagrees with the cosmological results but agrees better with the idealized simulations which essentially make the same implicit assumptions.

In contrast, in the isothermal model, the target CDM halo at $z=0$ is {\it not} treated as an {\it initial condition}, but instead used for the {\it boundary condition} at $r_1$.
In the context of trying to understand why distinct CDM haloes all have the universal NFW shape, it has been well-established that there is a correspondence between the density-profile shape and the shape of the mass assembly history \citep[e.g.,][]{Ludlow13}.
In this regard, using the target NFW profile to set the boundary condition means that we have implicitly used some information of the cosmological mass assembly history of the halo.
It is therefore reasonable to expect agreement with the cosmological simulations. 

It is still remarkable that the simplistic stitching at $r_1$ results in this high level of agreement and we caution against over-interpretating it physically. 
But we have verified that altering the detailed definition of $r_1$ does not change the qualitative agreement. 
For example, multiplying a constant factor in \eq{r1} will not change the overall shape of the $\rhot_0(\ttilde)$ track.
This implies that, when the isothermal assumption is valid, there is no hysteresis of the core. 
After all, the isothermal state is a thermodynamic equilibrium, so does not depend on how the state was reached.
  
We also caution that the comparison in \fig{IsothermalVersusGravothermal} should not be interpreted as a criticism of the fluid model, but instead a clarification of what it does as implemented in the literature \citep{LE80, Koda11, Balberg02, Pollack15, Essig19, Nishikawa20}.
To adapt it for better cosmological usage, one may want to explore revisions of the sort of the following. 
In particular, in order to model the SIDM counterpart of a CDM halo at $z=0$, it is reasonable to adopt the CDM profile at $z=z_{\rm form}$ as the initial condition, where $z_{\rm form}$ is a characteristic formation redshift of the halo. 
Accordingly, the mass, density, and velocity dispersion in \eq{fluid} shall be updated according to cosmological average trends to account for the growth history of the halo.
For example, for each timestep, one can add a mass increment to each radius bin self-similarly according to the instantaneous density profile, where the sum of the mass depositions across all the bins is equal to the average mass growth in that timestep. 
There are well-established empirical mass assembly histories and mass-concentration-redshift relations from CDM simulations \citep[e.g.,][]{Mcbride09,DM14}. 
We explore improvements of this sort in a future study (Yang et al. in preparation).

%---------------------------------------
\subsection{Gravothermal core-collapse and facilitation by the inhabitant galaxy}\label{sec:GC}

\begin{figure}
\includegraphics[width=0.46\textwidth]{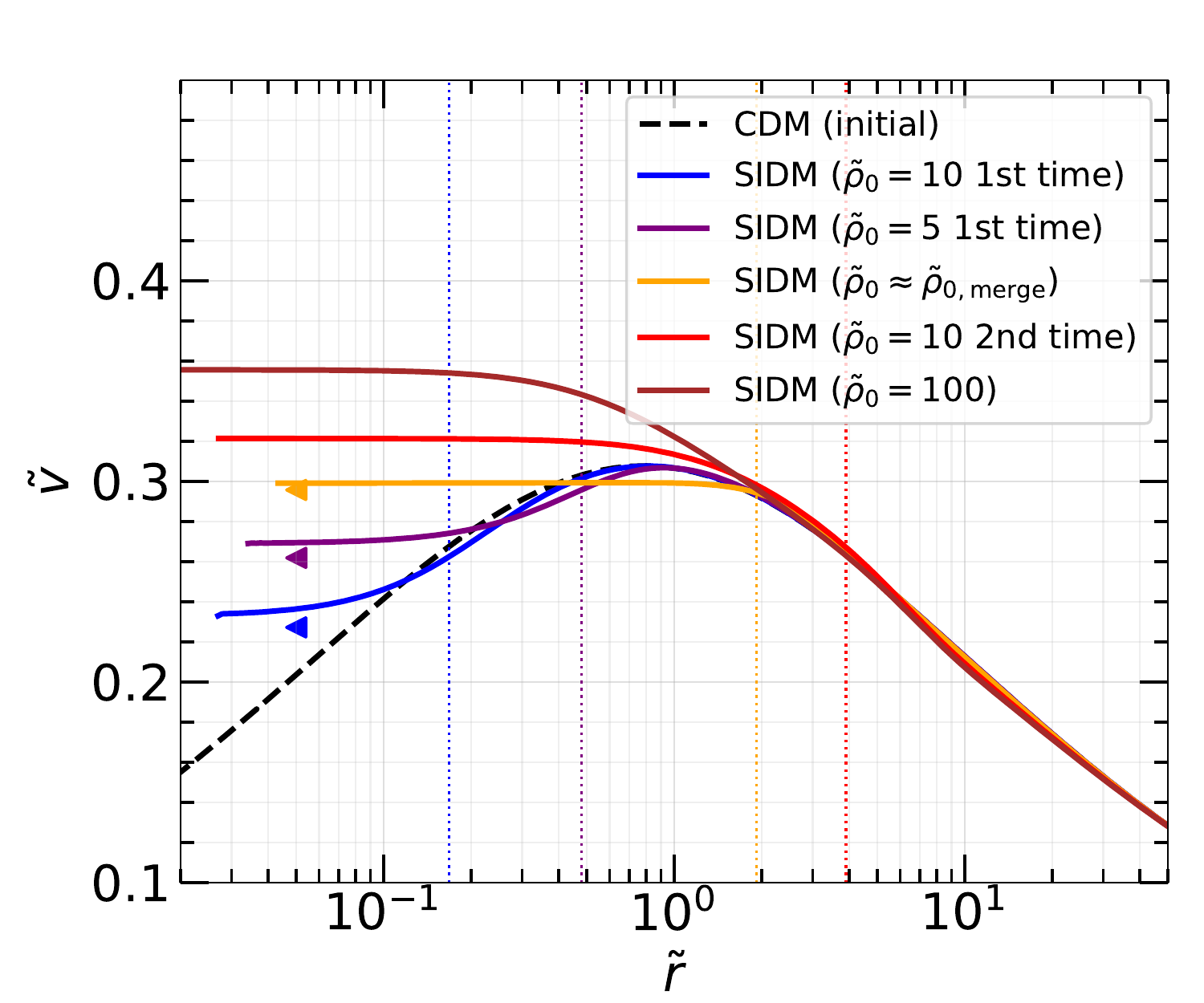}
\caption{
Evolution of the velocity dispersion profile from the fluid model. 
The dashed black line represents the initial CDM profile. 
The coloured solid lines represent the SIDM profiles at different times, as indicated. 
Notably, the orange line represents the result when $\rhot_0$ reaches the value when the low-density and the high-density solutions merge in the isothermal Jeans model, $\rhot_{\rm 0,merge}$.
Since the central density $\rhot_0$ initially decreases and later turns around, it will reach certain values twice -- e.g., both the blue and red lines here correspond to $\rhot_0=10$, but the former is during core-formation while the latter is during core collapse. 
The dotted vertical lines indicate the corresponding $r_1$ radii.
The triangles indicate the isothermal Jeans solutions of the central velocity dispersions (only for the first three cases, because the isothermal model is no longer applicable to the later stages of evolution).   
Obviously, beyond $\rhot_{\rm 0,merge}$, a negative velocity-dispersion gradient develops at $r\sim r_1$, core-collapse speeds up, and therefore the isothermal model stops working as the assumption of constant velocity dispersion at $r<r_1$ is no longer valid.
}
\label{fig:WhenIsothermalBreaks}
\end{figure}

As \fig{IsothermalVersusGravothermal} shows, gravothermal core-collapse occurs, i.e., the central density starts to increase, at $\ttilde\ga100$. 
Soon after the onset of core-collapse, one can see with the fluid model that the central velocity dispersion increases to a level that is higher than the CDM $v(r_1)$, and thus a steep negative velocity-dispersion gradient occurs at $r\la r_1$, as shown in \fig{WhenIsothermalBreaks}. 
Then, the flat isothermal core becomes significantly smaller than $r_1$ and thus gravothermal core-collapse speeds up.
Recall that the key assumption for the isothermal Jeans model is that the core has constant $v=v_0$ throughout the region $r<r_1$. 
This assumption holds at the onset of core-collapse, which is why the isothermal model is still able to capture the upturn in the central density. 
But the isothermal method fails as core-collapse continues, because $r_1$, as defined in \eq{r1}, increases with time, and therefore the assumption of $v=v_0$ within $r_1$ breaks when $v_0$ increases to be significantly higher than the peak of the CDM $v(r)$ profile. 

Recall that in the isothermal model, we accept the lower-density solution because realistic haloes form with properties closer to it. However, we emphasize that both the low-density and high-density solutions are physical, as they satisfy the Jeans-Poisson equation with constant velocity dispersion within $r_1$. 
A smooth transition between them is achieved shortly after the central density starts to increase, manifesting core-collapse. Therefore, we can practically use the moment when the two solutions merge as an indicator of gravothermal core-collapse. 
%Denoting this time as $t_{\rm merge}$, we find that the mirroring of the high-density solution with respect to $t_{\rm merge}$ gives a smooth extrapolation of the low-density solution.
%This is shown with the dashed orange curve in \fig{IsothermalVersusGravothermal}.
%The time of completing the core-collapse can therefore be phenomenologically approximated by $t_{\rm GC}\approx 2 t_{\rm merge}$, i.e., $\ttilde\approx 920$. 

The inhabitant galaxy facilitates core-collapse by making the halo contract in the first place. 
Naturally, this effect is particularly strong when the galaxy is massive and compact. 
To illustrate this, we highlight the region of core-collapse in the $\rhalf/\Rv-\Mb/\Mv$ space in \fig{HaloResponse}.
The operational definition of this region is that: for galaxies on the border of this region, the SIDM haloes that formed $10\Gyr$ ago have started core-collapse, such that no isothermal solution exists that joins smoothly the CDM outskirt (with $\delta^2<0.01$). 

\begin{figure}
\includegraphics[width=0.46\textwidth]{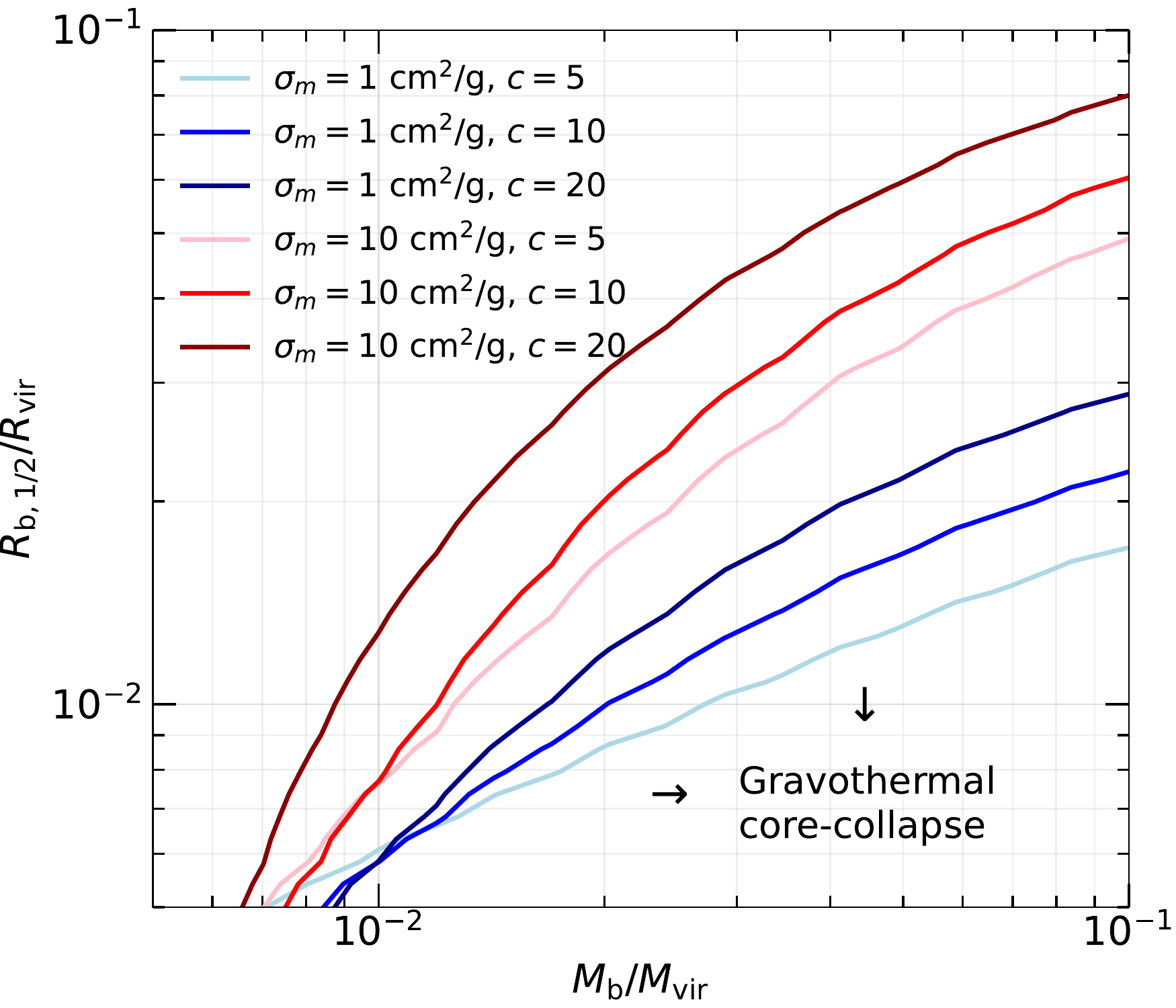}
\caption{Critical baryonic properties for gravothermal core-collapse -- the same as the boundary of the core-collapse region in \fig{HaloResponse}, but for different halo concentrations and SIDM cross sections, as indicated. The way to read this is: if the inhabitant galaxy has higher mass or more compact size than this threshold, the host SIDM halo formed $\sim10\Gyr$ ago will start core-collapse. Practically, systems sitting on the threshold have the two isothermal solutions merged into one. Beyond this threshold to the lower right corner, a smooth joint with $\delta^2<0.01$ between the SIDM core and the CDM-like outskirt is no longer achievable, or equivalently speaking, it is no longer possible to have constant velocity dispersion within $r_1$. See \se{GC} for details. }
\label{fig:GCthreshold}
\end{figure}

The region for core-collapse depends on the target CDM concentration and the self-interacting cross section, and becomes larger for higher $c$ and $\sigmam$.
This is illustrated in \fig{GCthreshold}. 
For instance, at $\Mb/\Mv=0.02$ fixed, haloes with $\rhalf/\Rv\la0.01$ will collapse if $\sigmam=1\cm^2/\g$.
The exact size limit is slightly lower for lower concentration. 
The galaxy-size limit becomes $\rhalf/\Rv\la0.02$ for $\sigmam=10\cm^2/\g$.
Given that numerous galaxies populate the region $\rhalf/\Rv\sim0.01$-$0.02$ observationally, this parameter space may potentially provide useful constraints on SIDM models. 
To this end, however, we think that it still requires more detailed understanding of how the inhabitant galaxies react to core-collapse.   
Regardless, the baryon-facilitated core-collapse itself might be a viable way to create compact bright dwarfs, which are common in the real Universe but are difficult to produce in cosmological hydro-simulations.  

For dwarf galaxies with $\Mb/\Mv\ll0.01$, there is basically no constraint on how compact the galaxy can get before core-collapse kicks in, as long as $c$ and $\sigmam$ are not extremely high.  

In short, with the current implementation of the isothermal Jeans model, although we cannot self-consistently describe core-collapse, we can phenomenologically delineate the onset of core-collapse as a function of the baryonic properties, given the target halo concentration and the cross section. 

%---------------------------------------------------------------------------------------------------------
\section{Conclusion}
\label{sec:conclusion}

In this paper, we combine the isothermal Jeans model and the prescription for adiabatic halo contraction into a fast semi-analytic procedure for calculating the density profile of SIDM haloes. 
This method takes the inputs of 1) a target CDM halo described by an NFW profile, and 2) an observable baryon distribution described by a Hernquist profile.
It computes the contraction, fits the contracted CDM halo with a Dekel-Zhao profile, and stitches an isothermal core to the CDM outskirt at the characteristic radius $r_1$ by minimizing the fractional difference in density and enclosed mass.
We have shown that this model works remarkably well compared to cosmological SIDM simulations both in dark-matter-only setups (Pippin) and with hydrodynamics and star formation (FIRE2-SIDM).
We provide a simple \coreNFW approximation formula for the dark-matter-only cases, where the characteristic core size of $\rc=0.45r_1$ universally applies to a wide range of cross sections and target CDM halo concentrations.

We use this model to study the response of SIDM haloes to their inhabitant galaxies.
We show that the halo response to the baryonic potential is more intensified and more diverse in SIDM than in CDM.
Notably, depending on the compactness of the baryonic distribution, the central dark-matter density slope can be cored, equally cuspy, or cuspier than the CDM counterpart -- a desirable feature in the context of the structural diversity of bright dwarf galaxies.  
We note that, while the model does not capture feedback-driven halo expansion and only considers adiabatic contraction, it agrees well with the FIRE2-SIDM simulations which incorporate both effects.  
We therefore argue that the dominant baryonic effect in the context of SIDM is adiabatic contraction, and that the details of baryonic feedback may be unimportant in SIDM models. 

The fast speed of the numerical implementation of the model enables the following analyses that would be otherwise challenging for numerical simulations.
We quantify the SIDM halo response on a fine mesh grid spanned by the baryon-to-total mass ratio $\Mb/\Mv$ and the ratio between the half mass radius and the virial-radius $\rhalf/\Rv$, in terms of the central logarithmic density slope, $s\equiv\rmd\ln\rho/\rmd\ln r|_{1\kpc}$, as well as the core density in units of the scale density of the reference CDM halo, $\rho_0/\rhos$.
With this, we are able to confirm with unprecedented precision that for typical Milky-Way-like hosts, the SIDM profiles are similar to their CDM counterparts -- an assumption often used in semi-analytic or idealized studies of SIDM satellite galaxies.

We also delineate the regime of gravothermal core-collapse in the space of galaxy mass versus galaxy size, $\Mb/\Mv-\rhalf/\Rv$.
This can be done for any choice of the cross section and the target CDM halo concentration.
For any given baryon-to-total ratio, there is a limit on how compact the galaxy can get in terms of $\rhalf/\Rv$, beyond which core-collapse will be triggered within the Hubble time. 
This threshold is lower (i.e., galaxies can be more compact) if the target CDM halo concentration is smaller or if the cross section is smaller. 
With $c=10$ and $\sigmam=1\cm^2/\g$, galaxy sizes cannot be smaller than $\sim0.01\Rv$ for typical baryon-to-total ratios of $\sim0.02$.
Given that numerous galaxies have  $\rhalf\sim0.01\Rv$, we think that this baryon-facilitated gravothermal core-collapse may provide useful constraints on SIDM models, if we can better understand how galaxies react to core-collapse.

Finally, we compare the isothermal Jeans model with the more sophisticated gravothermal fluid model which is extensively studied in the literature. 
We show that the isothermal model agrees better with cosmological simulations: they both show a steeper central-density decrease in the isothermal coring regime and a later gravothermal core-collapse compared to the fluid model.
On the contrary, the fluid model agrees well with idealized simulations of isolated haloes initialized with NFW profiles.
We argue that the difference originates from whether the target CDM profile is used for the {\it boundary condition} (as in the case of the isothermal model) or as the {\it initial condition} (as in the case of the fluid model). 

We have made our programs publicly available, including the programs for computing the profiles of SIDM haloes with baryons, as well as the programs that calculate the threshold for gravothermal core-collapse in the $\Mb/\Mv-\rhalf/\Rv$ space. They can be downloaded at \href{https://github.com/JiangFangzhou/SIDM}{https://github.com/JiangFangzhou/SIDM}. 
While we stick to Hernquist galaxies in the paper for self-consistency (as \eq{JeansPoisson} is based on Hernquist galaxies), the adiabatic contraction model of \citet{Gnedin04} actually also accommodates exponential disks and is implemented in the code.

%---------------------------------------
\section*{Acknowledgements}
We thank Ethan Nadler, Maya Silverman, Igor Palubski, and Dylan Folsom for helpful general discussions. 
%We thank Omid Sameie for helpful discussions on the FIRE2-SIDM simulations.
FJ is partially supported by the Troesh Scholarship from the California Institute of Technology.
AB, AHGP, ZCZ, and XD are supported in part by the NASA Astrophysics Theory Program under grant 80NSSC18K1014. 
ML and OS are supported by the DOE under Award Number DE-SC0007968 and the Binational Science Foundation (grant No. 2018140). 

%%%%%%%%%%%%%%%%%%%%%%%%%%%%%%%%%%%%%%%%%%%%%%%%%%

%%%%%%%%%%%%%%%%%% %% REFERENCES %%%%%%%%%%%%%%%%%%


\bibliographystyle{mnras}
\bibliography{SIDM} 

\begin{thebibliography}{}
\makeatletter
\relax
\def\mn@urlcharsother{\let\do\@makeother \do\$\do\&\do\#\do\^\do\_\do\%\do\~}
\def\mn@doi{\begingroup\mn@urlcharsother \@ifnextchar [ {\mn@doi@}
  {\mn@doi@[]}}
\def\mn@doi@[#1]#2{\def\@tempa{#1}\ifx\@tempa\@empty \href
  {http://dx.doi.org/#2} {doi:#2}\else \href {http://dx.doi.org/#2} {#1}\fi
  \endgroup}
\def\mn@eprint#1#2{\mn@eprint@#1:#2::\@nil}
\def\mn@eprint@arXiv#1{\href {http://arxiv.org/abs/#1} {{\tt arXiv:#1}}}
\def\mn@eprint@dblp#1{\href {http://dblp.uni-trier.de/rec/bibtex/#1.xml}
  {dblp:#1}}
\def\mn@eprint@#1:#2:#3:#4\@nil{\def\@tempa {#1}\def\@tempb {#2}\def\@tempc
  {#3}\ifx \@tempc \@empty \let \@tempc \@tempb \let \@tempb \@tempa \fi \ifx
  \@tempb \@empty \def\@tempb {arXiv}\fi \@ifundefined
  {mn@eprint@\@tempb}{\@tempb:\@tempc}{\expandafter \expandafter \csname
  mn@eprint@\@tempb\endcsname \expandafter{\@tempc}}}

\bibitem[\protect\citeauthoryear{Balberg \& Shapiro}{Balberg \&
  Shapiro}{2002}]{Balberg02}
Balberg S.,  Shapiro S.~L.,  2002, \mn@doi [Physical Review Letters]
  {10.1103/physrevlett.88.101301}, 88, 101301

\bibitem[\protect\citeauthoryear{Benson}{Benson}{2012}]{Benson12}
Benson A.~J.,  2012, \mn@doi [New Astronomy] {10.1016/j.newast.2011.07.004},
  17, 175

\bibitem[\protect\citeauthoryear{Blok, Walter, Brinks, Trachternach, Oh  \&
  Kennicutt}{Blok et~al.}{2008}]{deBlok08}
Blok W. J. G.~d.,  Walter F.,  Brinks E.,  Trachternach C.,  Oh S.-H.,
  Kennicutt R.~C.,  2008, \mn@doi [The Astronomical Journal]
  {10.1088/0004-6256/136/6/2648}, 136, 2648

\bibitem[\protect\citeauthoryear{Blumenthal, Faber, Flores  \&
  Primack}{Blumenthal et~al.}{1986}]{Blumenthal86}
Blumenthal G.~R.,  Faber S.~M.,  Flores R.,   Primack J.~R.,  1986, \mn@doi
  [The Astrophysical Journal] {10.1086/163867}, 301, 27

\bibitem[\protect\citeauthoryear{Bose et~al.,}{Bose et~al.}{2019}]{Bose19}
Bose S.,  et~al., 2019, \mn@doi [Monthly Notices of the Royal Astronomical
  Society] {10.1093/mnras/stz1168}

\bibitem[\protect\citeauthoryear{Chilingarian \& Zolotukhin}{Chilingarian \&
  Zolotukhin}{2015}]{Chilingarian15}
Chilingarian I.,  Zolotukhin I.,  2015, \mn@doi [Science]
  {10.1126/science.aaa3344}, 348, 418

\bibitem[\protect\citeauthoryear{Colin, Avila‐Reese, Valenzuela  \&
  Firmani}{Colin et~al.}{2002}]{Colin02}
Colin P.,  Avila‐Reese V.,  Valenzuela O.,   Firmani C.,  2002, \mn@doi [The
  Astrophysical Journal] {10.1086/344259}, 581, 777

\bibitem[\protect\citeauthoryear{Creasey, Sameie, Sales, Yu, Vogelsberger  \&
  Zavala}{Creasey et~al.}{2017}]{Creasey17}
Creasey P.,  Sameie O.,  Sales L.~V.,  Yu H.-B.,  Vogelsberger M.,   Zavala J.,
   2017, \mn@doi [Monthly Notices of the Royal Astronomical Society]
  {10.1093/mnras/stx522}, 468, 2283

\bibitem[\protect\citeauthoryear{Cruz et~al.,}{Cruz et~al.}{2021}]{Cruz21}
Cruz A.,  et~al., 2021, \mn@doi [Monthly Notices of the Royal Astronomical
  Society] {10.1093/mnras/staa3389}, 500, 2177

\bibitem[\protect\citeauthoryear{Dutton \& Macciò}{Dutton \&
  Macciò}{2014}]{DM14}
Dutton A.~A.,  Macciò A.~V.,  2014, \mn@doi [Monthly Notices of the Royal
  Astronomical Society] {10.1093/mnras/stu742}, 441, 3359

\bibitem[\protect\citeauthoryear{Elbert, Bullock, Garrison-Kimmel, Rocha,
  Oñorbe  \& Peter}{Elbert et~al.}{2015}]{Elbert15}
Elbert O.~D.,  Bullock J.~S.,  Garrison-Kimmel S.,  Rocha M.,  Oñorbe J.,
  Peter A. H.~G.,  2015, \mn@doi [Monthly Notices of the Royal Astronomical
  Society] {10.1093/mnras/stv1470}, 453, 29

\bibitem[\protect\citeauthoryear{Elbert, Bullock, Kaplinghat, Garrison-Kimmel,
  Graus  \& Rocha}{Elbert et~al.}{2018}]{Elbert18}
Elbert O.~D.,  Bullock J.~S.,  Kaplinghat M.,  Garrison-Kimmel S.,  Graus
  A.~S.,   Rocha M.,  2018, \mn@doi [The Astrophysical Journal]
  {10.3847/1538-4357/aa9710}, 853, 109

\bibitem[\protect\citeauthoryear{Essig, McDermott, Yu  \& Zhong}{Essig
  et~al.}{2019}]{Essig19}
Essig R.,  McDermott S.~D.,  Yu H.-B.,   Zhong Y.-M.,  2019, \mn@doi [Physical
  Review Letters] {10.1103/physrevlett.123.121102}, 123, 121102

\bibitem[\protect\citeauthoryear{Freundlich et~al.,}{Freundlich
  et~al.}{2020}]{Freundlich20}
Freundlich J.,  et~al., 2020, MNRAS, 499, 2912

\bibitem[\protect\citeauthoryear{Gnedin, Kravtsov, Klypin  \& Nagai}{Gnedin
  et~al.}{2004}]{Gnedin04}
Gnedin O.~Y.,  Kravtsov A.~V.,  Klypin A.~A.,   Nagai D.,  2004, \mn@doi [The
  Astrophysical Journal] {10.1086/424914}, 616, 16

\bibitem[\protect\citeauthoryear{Gnedin, Ceverino, Gnedin, Klypin, Kravtsov,
  Levine, Nagai  \& Yepes}{Gnedin et~al.}{2011}]{Gnedin11}
Gnedin O.~Y.,  Ceverino D.,  Gnedin N.~Y.,  Klypin A.~A.,  Kravtsov A.~V.,
  Levine R.,  Nagai D.,   Yepes G.,  2011, arXiv:1108.5736

\bibitem[\protect\citeauthoryear{Jiang, Dekel, Freundlich, Romanowsky, Dutton,
  Macciò  \& Cintio}{Jiang et~al.}{2019}]{Jiang19}
Jiang F.,  Dekel A.,  Freundlich J.,  Romanowsky A.~J.,  Dutton A.~A.,  Macciò
  A.~V.,   Cintio A.~D.,  2019, \mn@doi [Monthly Notices of the Royal
  Astronomical Society] {10.1093/mnras/stz1499}, 487, 5272

\bibitem[\protect\citeauthoryear{Jiang, Dekel, Freundlich, van den Bosch,
  Green, Hopkins, Benson  \& Du}{Jiang et~al.}{2021}]{Jiang21}
Jiang F.,  Dekel A.,  Freundlich J.,  van den Bosch F.~C.,  Green S.~B.,
  Hopkins P.~F.,  Benson A.,   Du X.,  2021, \mn@doi [Monthly Notices of the
  Royal Astronomical Society] {10.1093/mnras/staa4034}, 502, 621

\bibitem[\protect\citeauthoryear{Jiang, Kaplinghat, Lisanti  \& Slone}{Jiang
  et~al.}{2022}]{Jiang22}
Jiang F.,  Kaplinghat M.,  Lisanti M.,   Slone O.,  2022, arXiv: 2108.03243

\bibitem[\protect\citeauthoryear{Kamada, Kaplinghat, Pace  \& Yu}{Kamada
  et~al.}{2017}]{Kamada17}
Kamada A.,  Kaplinghat M.,  Pace A.~B.,   Yu H.-B.,  2017, \mn@doi [Physical
  Review Letters] {10.1103/physrevlett.119.111102}, 119, 111102

\bibitem[\protect\citeauthoryear{Kaplinghat, Keeley, Linden  \& Yu}{Kaplinghat
  et~al.}{2014}]{Kaplinghat14}
Kaplinghat M.,  Keeley R.~E.,  Linden T.,   Yu H.-B.,  2014, \mn@doi [Physical
  Review Letters] {10.1103/physrevlett.113.021302}, 113, 021302

\bibitem[\protect\citeauthoryear{Kaplinghat, Tulin  \& Yu}{Kaplinghat
  et~al.}{2016}]{Kaplinghat16}
Kaplinghat M.,  Tulin S.,   Yu H.-B.,  2016, \mn@doi [Physical Review Letters]
  {10.1103/physrevlett.116.041302}, 116, 041302

\bibitem[\protect\citeauthoryear{Kaplinghat, Ren  \& Yu}{Kaplinghat
  et~al.}{2020}]{Kaplinghat20}
Kaplinghat M.,  Ren T.,   Yu H.-B.,  2020, \mn@doi [Journal of Cosmology and
  Astroparticle Physics] {10.1088/1475-7516/2020/06/027}, 2020, 027

\bibitem[\protect\citeauthoryear{Kochanek \& White}{Kochanek \&
  White}{2000}]{Kochanek00}
Kochanek C.~S.,  White M.,  2000, \mn@doi [The Astrophysical Journal]
  {10.1086/317149}, 543, 514

\bibitem[\protect\citeauthoryear{Koda \& Shapiro}{Koda \&
  Shapiro}{2011}]{Koda11}
Koda J.,  Shapiro P.~R.,  2011, \mn@doi [Monthly Notices of the Royal
  Astronomical Society] {10.1111/j.1365-2966.2011.18684.x}, 415, 1125

\bibitem[\protect\citeauthoryear{Koda, Yagi, Yamanoi  \& Komiyama}{Koda
  et~al.}{2015}]{Koda15}
Koda J.,  Yagi M.,  Yamanoi H.,   Komiyama Y.,  2015, \mn@doi [The
  Astrophysical Journal Letters] {10.1088/2041-8205/807/1/l2}, 807, L2

\bibitem[\protect\citeauthoryear{Komatsu et~al.,}{Komatsu
  et~al.}{2011}]{Komatsu11}
Komatsu E.,  et~al., 2011, \mn@doi [The Astrophysical Journal Supplement]
  {10.1088/0067-0049/192/2/18}, 192

\bibitem[\protect\citeauthoryear{Lelli, McGaugh  \& Schombert}{Lelli
  et~al.}{2016}]{Lelli16}
Lelli F.,  McGaugh S.~S.,   Schombert J.~M.,  2016, \mn@doi [The Astronomical
  Journal] {10.3847/0004-6256/152/6/157}, 152, 157

\bibitem[\protect\citeauthoryear{Ludlow et~al.,}{Ludlow
  et~al.}{2013}]{Ludlow13}
Ludlow A.~D.,  et~al., 2013, \mn@doi [Monthly Notices of the Royal Astronomical
  Society] {10.1093/mnras/stt526}, 432, 1103

\bibitem[\protect\citeauthoryear{Lynden-Bell \& Eggleton}{Lynden-Bell \&
  Eggleton}{1980}]{LE80}
Lynden-Bell D.,  Eggleton P.~P.,  1980, \mn@doi [Monthly Notices of the Royal
  Astronomical Society] {10.1093/mnras/191.3.483}, 191, 483

\bibitem[\protect\citeauthoryear{McBride, Fakhouri  \& Ma}{McBride
  et~al.}{2009}]{Mcbride09}
McBride J.,  Fakhouri O.,   Ma C.-P.,  2009, \mn@doi [Monthly Notices of the
  Royal Astronomical Society] {10.1111/j.1365-2966.2009.15329.x}, 398, 1858

\bibitem[\protect\citeauthoryear{Moster, Naab  \& White}{Moster
  et~al.}{2013}]{Moster13}
Moster B.~P.,  Naab T.,   White S. D.~M.,  2013, \mn@doi [Monthly Notices of
  the Royal Astronomical Society] {10.1093/mnras/sts261}, 428, 3121

\bibitem[\protect\citeauthoryear{Nadler, Banerjee, Adhikari, Mao  \&
  Wechsler}{Nadler et~al.}{2020}]{Nadler20}
Nadler E.~O.,  Banerjee A.,  Adhikari S.,  Mao Y.-Y.,   Wechsler R.~H.,  2020,
  \mn@doi [The Astrophysical Journal] {10.3847/1538-4357/ab94b0}, 896, 112

\bibitem[\protect\citeauthoryear{Navarro, Frenk  \& White}{Navarro
  et~al.}{1997}]{NFW97}
Navarro J.~F.,  Frenk C.~S.,   White S. D.~M.,  1997, \mn@doi [The
  Astrophysical Journal] {10.1086/304888}, 490, 493

\bibitem[\protect\citeauthoryear{Nishikawa, Boddy  \& Kaplinghat}{Nishikawa
  et~al.}{2020}]{Nishikawa20}
Nishikawa H.,  Boddy K.~K.,   Kaplinghat M.,  2020, \mn@doi [Physical Review D]
  {10.1103/physrevd.101.063009}, 101, 063009

\bibitem[\protect\citeauthoryear{Oh et~al.,}{Oh et~al.}{2015}]{Oh15}
Oh S.-H.,  et~al., 2015, \mn@doi [The Astronomical Journal]
  {10.1088/0004-6256/149/6/180}, 149, 180

\bibitem[\protect\citeauthoryear{Peter, Rocha, Bullock  \& Kaplinghat}{Peter
  et~al.}{2013}]{Peter13}
Peter A. H.~G.,  Rocha M.,  Bullock J.~S.,   Kaplinghat M.,  2013, \mn@doi
  [Monthly Notices of the Royal Astronomical Society] {10.1093/mnras/sts535},
  430, 105

\bibitem[\protect\citeauthoryear{Pollack, Spergel  \& Steinhardt}{Pollack
  et~al.}{2015}]{Pollack15}
Pollack J.,  Spergel D.~N.,   Steinhardt P.~J.,  2015, \mn@doi [The
  Astrophysical Journal] {10.1088/0004-637x/804/2/131}, 804, 131

\bibitem[\protect\citeauthoryear{Read, Agertz  \& Collins}{Read
  et~al.}{2016}]{Read16}
Read J.~I.,  Agertz O.,   Collins M. L.~M.,  2016, \mn@doi [Monthly Notices of
  the Royal Astronomical Society] {10.1093/mnras/stw713}, 459, 2573

\bibitem[\protect\citeauthoryear{Relatores et~al.,}{Relatores
  et~al.}{2019}]{Relatores19}
Relatores N.~C.,  et~al., 2019, \mn@doi [Monthly Notices of the Royal
  Astronomical Society] {10.3847/1538-4357/ab5305}, 887, 94

\bibitem[\protect\citeauthoryear{Ren, Kwa, Kaplinghat  \& Yu}{Ren
  et~al.}{2019}]{Ren19}
Ren T.,  Kwa A.,  Kaplinghat M.,   Yu H.-B.,  2019, \mn@doi [Physical Review X]
  {10.1103/PhysRevX.9.031020}, 9, 031020

\bibitem[\protect\citeauthoryear{Robertson, Harvey, Massey, Eke, McCarthy,
  Jauzac, Li  \& Schaye}{Robertson et~al.}{2019}]{Robertson19}
Robertson A.,  Harvey D.,  Massey R.,  Eke V.,  McCarthy I.~G.,  Jauzac M.,  Li
  B.,   Schaye J.,  2019, \mn@doi [Monthly Notices of the Royal Astronomical
  Society] {10.1093/mnras/stz1815}, 488, 3646

\bibitem[\protect\citeauthoryear{Robertson, Massey, Eke, Schaye  \&
  Theuns}{Robertson et~al.}{2021}]{Robertson21}
Robertson A.,  Massey R.,  Eke V.,  Schaye J.,   Theuns T.,  2021, \mn@doi
  [Monthly Notices of the Royal Astronomical Society] {10.1093/mnras/staa3954},
  501

\bibitem[\protect\citeauthoryear{Rocha, Peter, Bullock, Kaplinghat,
  Garrison-Kimmel, Oñorbe  \& Moustakas}{Rocha et~al.}{2013}]{Rocha13}
Rocha M.,  Peter A. H.~G.,  Bullock J.~S.,  Kaplinghat M.,  Garrison-Kimmel S.,
   Oñorbe J.,   Moustakas L.~A.,  2013, \mn@doi [Monthly Notices of the Royal
  Astronomical Society] {10.1093/mnras/sts514}, 430, 81

\bibitem[\protect\citeauthoryear{Sagunski, Gad-Nasr, Colquhoun, Robertson  \&
  Tulin}{Sagunski et~al.}{2021}]{Sagunski21}
Sagunski L.,  Gad-Nasr S.,  Colquhoun B.,  Robertson A.,   Tulin S.,  2021,
  \mn@doi [Journal of Cosmology and Astroparticle Physics]
  {10.1088/1475-7516/2021/01/024}, 2021, 024

\bibitem[\protect\citeauthoryear{Sameie et~al.,}{Sameie
  et~al.}{2021}]{Sameie21}
Sameie O.,  et~al., 2021, \mn@doi [Monthly Notices of the Royal Astronomical
  Society] {10.1093/mnras/stab2173}, 507, 720

\bibitem[\protect\citeauthoryear{Shen, Hopkins, Necib, Jiang, Boylan-Kolchin
  \& Wetzel}{Shen et~al.}{2021}]{Shen21}
Shen X.,  Hopkins P.~F.,  Necib L.,  Jiang F.,  Boylan-Kolchin M.,   Wetzel A.,
   2021, \mn@doi [Monthly Notices of the Royal Astronomical Society]
  {10.1093/mnras/stab2042}, 506, 4421

\bibitem[\protect\citeauthoryear{Shi, Zhang, Wang, Chen, Gu, Yu  \& Li}{Shi
  et~al.}{2021}]{Shi21}
Shi Y.,  Zhang Z.-Y.,  Wang J.,  Chen J.,  Gu Q.,  Yu X.,   Li S.,  2021,
  \mn@doi [The Astrophysical Journal] {10.3847/1538-4357/abd777}, 909, 20

\bibitem[\protect\citeauthoryear{Somerville et~al.,}{Somerville
  et~al.}{2018}]{Somerville18}
Somerville R.~S.,  et~al., 2018, \mn@doi [Monthly Notices of the Royal
  Astronomical Society] {10.1093/mnras/stx2040}, 473, 2714

\bibitem[\protect\citeauthoryear{Springel}{Springel}{2010}]{Springel10}
Springel V.,  2010, \mn@doi [Monthly Notices of the Royal Astronomical Society]
  {10.1111/j.1365-2966.2009.15715.x}, 401, 791

\bibitem[\protect\citeauthoryear{Vogelsberger, Zavala  \& Loeb}{Vogelsberger
  et~al.}{2012}]{Vogelsberger12}
Vogelsberger M.,  Zavala J.,   Loeb A.,  2012, \mn@doi [Monthly Notices of the
  Royal Astronomical Society] {10.1111/j.1365-2966.2012.21182.x}, 423, 3740

\bibitem[\protect\citeauthoryear{Weinberger, Springel  \& Pakmor}{Weinberger
  et~al.}{2020}]{Weinberger20}
Weinberger R.,  Springel V.,   Pakmor R.,  2020, \mn@doi [The Astrophysical
  Journal Supplement Series] {10.3847/1538-4365/ab908c}, 248, 32

\bibitem[\protect\citeauthoryear{Zeng, Peter, Du, Benson, Kim, Jiang,
  Cyr-Racine  \& Vogelsberger}{Zeng et~al.}{2021}]{Zeng21}
Zeng Z.~C.,  Peter A. H.~G.,  Du X.,  Benson A.,  Kim S.,  Jiang F.,
  Cyr-Racine F.-Y.,   Vogelsberger M.,  2021, arXiv

\bibitem[\protect\citeauthoryear{Zentner, Dandavate, Slone  \& Lisanti}{Zentner
  et~al.}{2022}]{Zentner22}
Zentner A.,  Dandavate S.,  Slone O.,   Lisanti M.,  2022, arXiv

\bibitem[\protect\citeauthoryear{van~den Bosch \& Ogiya}{van~den Bosch \&
  Ogiya}{2018}]{vdB18}
van~den Bosch F.~C.,  Ogiya G.,  2018, \mn@doi [Monthly Notices of the Royal
  Astronomical Society] {10.1093/mnras/sty084}, 475, 4066

\makeatother
\end{thebibliography}

%%%%%%%%%%%%%%%%%%%%%%%%%%%%%%%%%%%%%%%%%%%%%%%%%%

%%%%%%%%%%%%%%%%% APPENDICES %%%%%%%%%%%%%%%%%%%%%

\appendix

%---------------------------------------------------------------------------------------------------------
\section{Two solutions of the Isothermal Jeans stitching}
\label{app:TwoSolutions}

\begin{figure*}	
	\includegraphics[width=\textwidth]{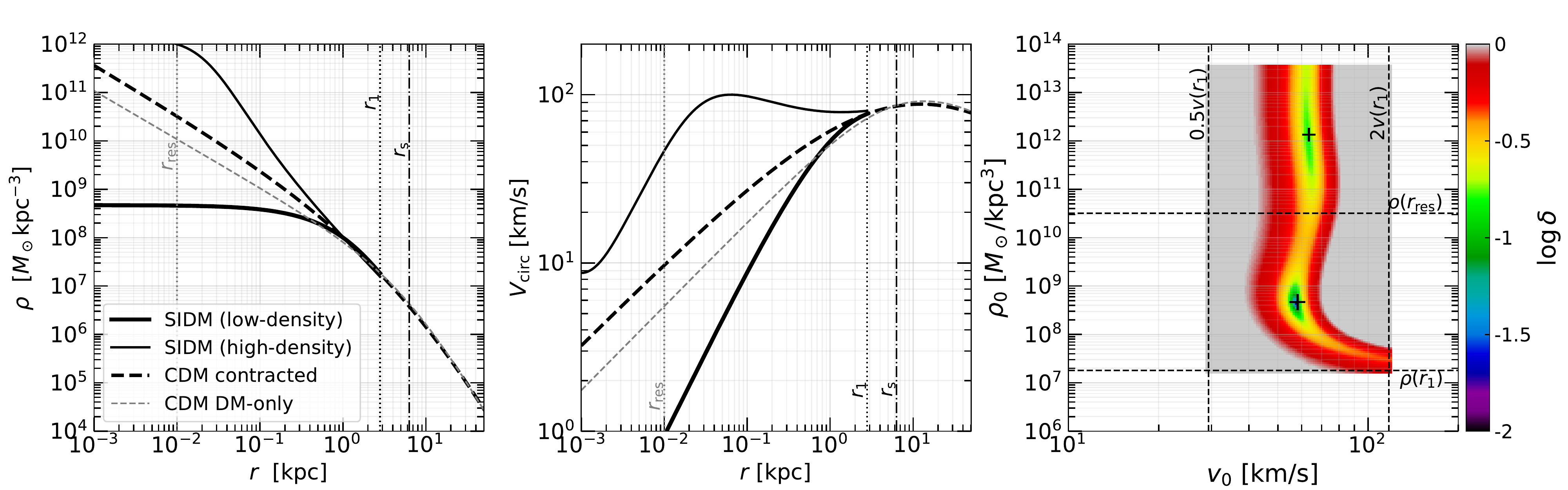}
	\includegraphics[width=\textwidth]{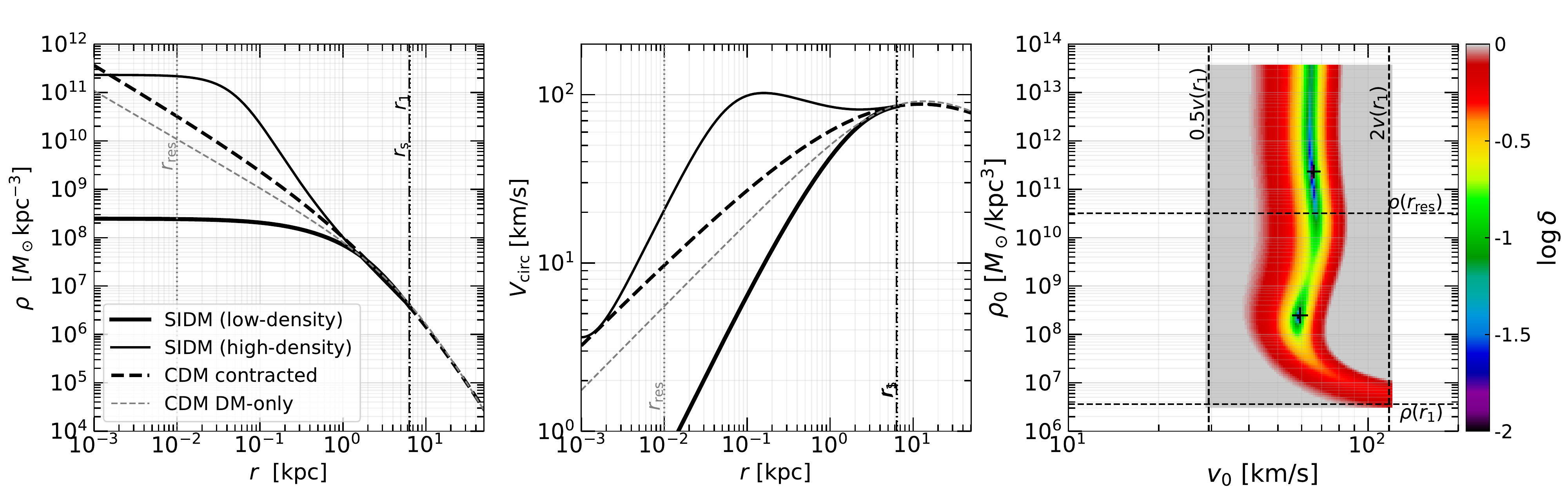}
	\includegraphics[width=\textwidth]{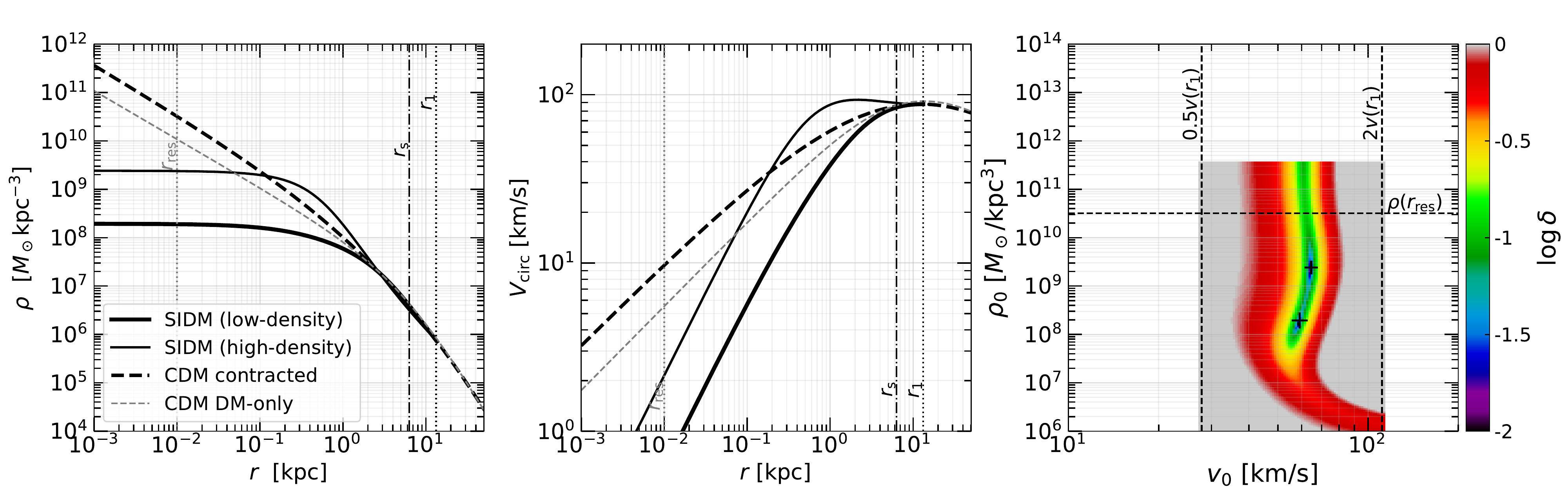}
	\includegraphics[width=\textwidth]{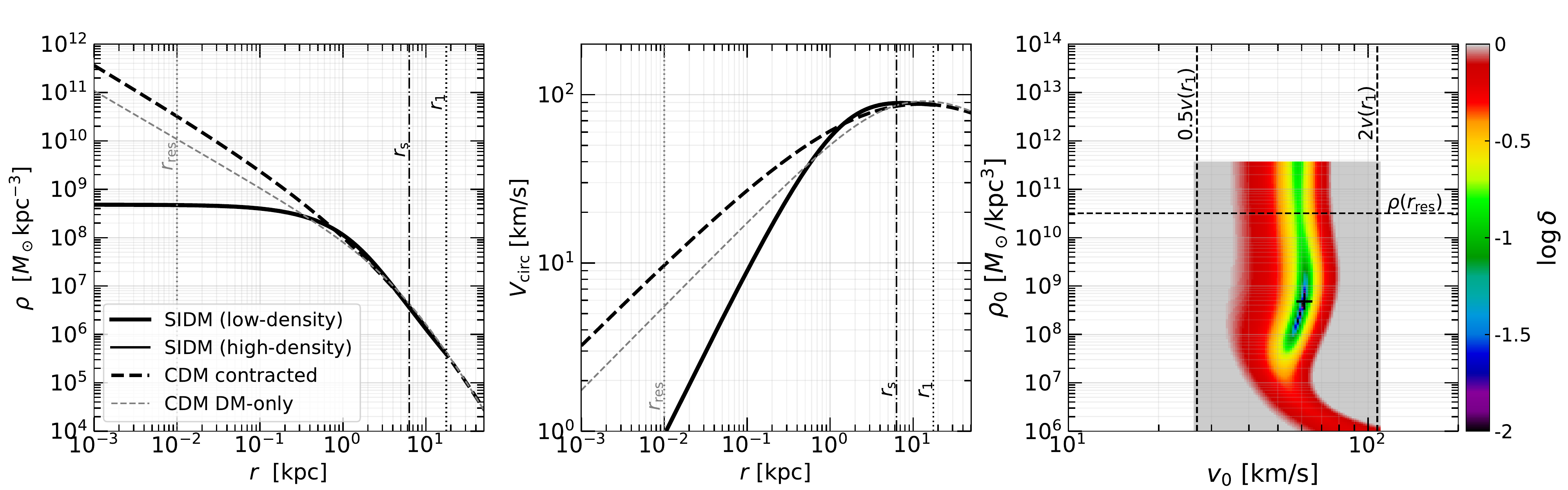}
    \caption{ 
    The same as \fig{stitching}, but for a series of different halo ages: $\tage=2, 10, 50$, and $100\Gyr$. Note that: 1) the central density $\rho_0$ decreases first at $\tage\la50\Gyr$ and then increases between $\tage=50\Gyr$ and $100\Gyr$; 2) the low-density and high-density solutions get closer as the system evolves and finally merge -- this is when the isothermal assumption starts to break, with $v_0$ reaching the highest value allowed by the isothermal assumption, and this is when gravothermal core-collapse kicks in. 
    }
    \label{fig:TwoSolutions}
\end{figure*}

In \se{model} and \fig{stitching}, we illustrated the workflow of the isothermal Jeans model and showed that there are two islands of minima of the `stitching error' at $r_1$, in the space of central dark-matter density $\rho_0$ and the core velocity dispersion $v_0$.
There, we showed an example of a system of $\tage=5\Gyr$, $\Mv=10^{11}\Msun$, $c=15$, $\Mb=10^{9}\Msun$, and $\rhalf=1.9\kpc$, for a cross section of $\sigmam=1\cm^2/\g$. 
Here, as shown in \fig{TwoSolutions}, we extend the exercise to a series of different halo ages, $\tage=2$, 10, 50, and 100 \Gyr, with everything else the same.
This effectively shows the evolution of the system.

As the system evolves, the two minima of $\delta^2$ first both decrease in $\rho_0$ ($\tage=2$ and 10\Gyr); then, the lower-density solution turns around ($\tage=50\Gyr$) and finally the two solutions merge ($\tage=100\Gyr$), marking the onset of gravothermal core-collapse. 

This trend actually holds as long as the system `evolves' in terms of the dimensionless time $\ttilde\equiv \tage/t_0 = 8\sqrt{G} \sigmam \rhos^{3/2}\rs\tage$, so it can be achieved also by increasing $\sigmam$ or $c$. For example, the central density track of the Pippin simulations as we showed in \fig{IsothermalVersusGravothermal} is obtained by increasing $\sigmam$ with everything else fixed.  

%---------------------------------------------------------------------------------------------------------
\section{Idealized simulations}\label{app:idealized}

For \se{gravothermal}, in addition to comparing with the published cosmological Pippin simulations, we also compared the models to idealized simulations of isolated SIDM haloes using the \texttt{Arepo} code \citep{Springel10, Weinberger20}. 
\texttt{Arepo} comes with a default module of dark matter self-interactions with the form of two-body scattering \citep{Vogelsberger12}.
This code is intensively used in the recent study of \citet{Zeng21} on SIDM subhaloes.
The initial conditions are generated with NFW profiles of $\Mv=10^{12}\Msun$ with a concentration parameter of $c=45$ or 90, using the code \texttt{SpherIC}. 
High concentration values are adopted to facilitate the gravothermal evolution. 
The particle mass is $10^7\Msun$.
The gravitational softening length $\epsilon$ of each halo is decided following the criteria of \citet{vdB18} such that:
\be
    \epsilon = r_s  f(c) \sqrt{\frac{0.32 (N_{\rm p}/1000)^{-0.8}}{1.12c^{1.26}}},
\ee
where $r_s$ is the scale radius of the initial NFW halo, $c$ is the initial concentration, $f(c)=\ln{(1+c)} - c/(1+c) $, and $N_{\rm p}$ is the number of simulation particles. 
The haloes are evolved with self-interaction cross section $\sigmam=10 \rm\ cm^2/g$ until a core is well developed in the centre. 
We emphasize that for the dimensionless $\rhot_0$-$\ttilde$ space (\fig{IsothermalVersusGravothermal}) in which we compare the results, the mass and the concentration of the halo or the cross section has little impact on the results. 
The central density $\rho_0$ is defined as the average density of the innermost 100 particles.

%%%%%%%%%%%%%%%%%%%%%%%%%%%%%%%%%%%%%%%%%%%%%%%%%%

% Don't change these lines
\bsp	% typesetting comment
\label{lastpage}
\end{document}